\documentclass[preprint,showpacs,preprintnumbers,amssymb,superscriptaddress,aps,prd,nofootinbib,11pt]{revtex4-1}

\selectfont

\usepackage{graphicx}       
\usepackage{dcolumn}        
\usepackage{bm}             
\usepackage{psfrag}
\usepackage{tensor}
\usepackage[usenames]{color}
\definecolor{navyblue}{rgb}{0.0, 0.0, 0.5}
\usepackage[linktocpage,colorlinks=true,allcolors=navyblue]{hyperref}
\usepackage{amsmath}
\usepackage{subfigure}

\newcommand{\be}{\begin{equation}}
\newcommand{\ee}{\end{equation}}

\newcommand{\cD}{\mathcal{D}}
\newcommand{\cL}{\mathcal{L}}
\newcommand{\cB}{\mathcal{B}}

\newcommand{\Sminus}{S_{-1}}
\newcommand{\Splus}{S_{+1}}

\newcommand{\bl}[1]{\textcolor{blue}{#1}}

\newcommand{\beq}{\begin{equation}}
\newcommand{\eeq}{\end{equation}}
\newcommand{\mbar}{\overline{m}}
\newcommand{\nn}{\nonumber}

\newcommand{\Pin}{P^{\ell m, h}} 
\newcommand{\Pup}{P^{\ell m, \infty}}
\newcommand{\Pinstatic}{P^{\ell 0, h}} 
\newcommand{\Pupstatic}{P^{\ell 0, \infty}}

\newcommand{\cF}{\mathbb{F}}

\begin{document}

\preprint{}

\title{Electromagnetic self-force on a charged particle on Kerr spacetime: \\ equatorial circular orbits}

\author{Theo Torres}
 \email{t.torres-vicente@sheffield.ac.uk}
\affiliation{Consortium for Fundamental Physics,
School of Mathematics and Statistics,
University of Sheffield, Hicks Building, Hounsfield Road, Sheffield S3 7RH, United Kingdom}

\author{Sam R. Dolan}
 \email{s.dolan@sheffield.ac.uk}
\affiliation{Consortium for Fundamental Physics,
School of Mathematics and Statistics,
University of Sheffield, Hicks Building, Hounsfield Road, Sheffield S3 7RH, United Kingdom}

\date{\today}

\begin{abstract}
We calculate the self-force acting on a charged particle on a circular geodesic orbit in the equatorial plane of a rotating black hole. We show by direct calculation that the dissipative self-force balances with the sum of the flux radiated to infinity and through the black hole horizon. Prograde orbits are found to stimulate black hole superradiance, but we confirm that the condition for floating orbits cannot be met. We calculate the conservative component of the self-force by application of the mode sum regularization method, and we present a selection of numerical results. By numerical fitting, we extract the leading-order coefficients in post-Newtonian expansions. The self-force on the innermost stable circular orbits of the Kerr spacetime is calculated, and comparisons are drawn between the electromagnetic and gravitational self forces.
\end{abstract}

\maketitle

\section{Introduction}

It is well-known that classical field theory is unable to satisfactorily account for the observed stability of the hydrogen atom. In the `planetary' version of the Rutherford atomic model \cite{rutherford1914lvii}, a point-like electron orbits the atomic nucleus. The centripetal acceleration of the charged electron generates electromagnetic (EM) radiation at the orbital frequency of $\sim 10^{15}$ Hz and, consequently, a radiation-reaction force acts upon the electron, causing the rapid collapse of the atom within $10^{-8}$\,s. Invoking the Abraham-Lorentz \cite{abraham1905theorie,Lorentz:1892} force law, 
\be
\mathbf{F} = \frac{2}{3} \frac{e^2}{4 \pi \epsilon_0 c^3} \dot{\mathbf{a}} , \label{eq:abraham-lorentz}
\ee
non-relativistic classical theory\footnote{For a fully relativistic treatment, one would instead start with the Abraham-Lorentz-Dirac equation \cite{Dirac:1938}; but note that $v/c < 0.01$ for a point-like electron at the Bohr radius.} implies that a point-like electron on a quasi-circular inspiral trajectory will generate EM radiation with the following `chirp' profile:
\be
f(t) \approx \frac{1}{4\pi \alpha} \sqrt{ \frac{c}{a_0}} (t_0 -t)^{-1/2} \, .  \label{eq:chirpEM}
\ee
Here $f$ is the EM frequency, $c$ is the speed of light, $\alpha$ is the fine-structure constant, $a_0$ is the Bohr radius, and $t_0$ is the time of collision (see Appendix \ref{appendix:Newtonian}).

There is no experimental support for collapsing atoms and/or EM chirps, of course. To the contrary,  experiments with electric discharges from the 1850s onwards show that atoms emit EM radiation at certain discrete frequencies \cite{Schawlow:1982spectroscopy}. Tension between theory and experiment led to the introduction of the Bohr-Rutherford atomic model \cite{Baily2013early}, and on to quantum theory itself. However, the idea of a continuous chirp from orbiting bodies has re-emerged as a key concept on a very different scale in the universe.

Compact binaries in astrophysics undergo an inspiral, due to the emission of gravitational waves. A pair of compact bodies of masses $M_1$ and $M_2$, on quasi-circular orbits about the centre of mass, will radiate gravitational waves predominantly in the quadrupole mode ($\ell = 2$) at twice the orbital frequency \cite{einstein1918gravitationswellen, Peters:1963}. 
Consequently, the binary system loses energy, and the GW frequency increases with a characteristic chirp profile,
\be
f(t) \approx \frac{5}{8 \pi} \left( \frac{5G \mathcal{M}}{c^3} \right)^{-5/8} (t_0 - t)^{-3/8}  ,  \label{eq:chirp}
\ee
where $\mathcal{M} = (M_1 M_2)^{3/5} / (M_1 + M_2)^{1/5}$ is the \emph{chirp mass} \cite{Abbott:2016bqf}. In 2017, the spectrogram of the gravitational wave signal from a binary neutron star inspiral was found to track this chirp profile remarkably closely over the last $\sim 100$ seconds before merger \cite{TheLIGOScientific:2017qsa}, despite the fact that, formally, Eq.~(\ref{eq:chirp}) arises only from the leading-order term of a post-Newtonian expansion for the radiated flux \cite{Peters:1963}.

In this article we consider the radiation-reaction process for a charged particle orbiting a black hole of mass $M$, rather than a charged nucleus. We shall assume that the length-scales of the particle, such as its Compton wavelength, are substantially smaller than the curvature scale, so that classical field theory provides an adequate framework. One might expect that, since the gravitational force and the Coulomb force both follow inverse square-laws in the weak-field, the radiation reaction process will proceed in a broadly similar fashion, producing a chirp frequency which scales with $(t_0 - t)^{-1/2}$ while $v \ll c$ and $r \gg GM/c^2$. However, an important difference that cannot be overlooked is that the spacetime of a black hole is curved, not flat.

The first expression for an EM self-force on a weakly curved spacetime was obtained by DeWitt-Morette and DeWitt \cite{DeWitt:1964} in 1964. The self-force on a particle of charge $q$ on a vacuum spacetime characterized by a Newtonian potential $\Phi_N = GM/c^2 r \ll 1$ is given by
\be
 \mathbf{F}_{\text{self}} \approx \frac{q^2}{4 \pi \epsilon_0 c^3}  \left( \frac{2}{3} \frac{d \mathbf{g}}{dt} + \frac{GMc}{r^3} \hat{\mathbf{r}} \right),  \label{eq:DeWitt}
\ee
where $\mathbf{g} = - c^{2} \boldsymbol{\nabla} \Phi_N$ is the Newtonian gravitational field. The first term in parantheses in Eq.~(\ref{eq:DeWitt}) is the standard Abraham-Lorentz force, which leads to the dissipation of orbital energy, and thus to an analogue of Eq.~(\ref{eq:chirpEM}). The second term is a \emph{conservative} correction to the Newtonian force $m \mathbf{g}$, which is \emph{not} present in flat spacetime. Analogous equations were obtained for scalar and gravitational self-forces in weakly-curved spacetimes in Ref.~\cite{Pfenning:2000zf}. 

To move beyond the Newtonian/weak-field context, we must acknowledge several key differences between a point mass in Newtonian theory and a black hole in general relativity. First, there exists an innermost stable circular orbit (ISCO), inside of which circular orbits cannot be sustained. Second, orbital velocities are sizable ($v/c \sim 0.4$ at the Schwarzschild ISCO), necessitating a fully relativistic description. Third, the issue of regularization is more subtle in a curved space-time, and Dirac's time-reversal approach (`half-advanced-minus-retarded') breaks down and requires modification \cite{DeWitt:1960fc, Detweiler:2002mi, Gralla:2009md}. 

The conservative component of the EM self-force leads to a shift in the orbital energy and angular momentum, and to a shift in the ISCO radius and frequency. The dissipative component of the EM self-force leads to orbital decay, and to the possibility of two interesting phenomena: floating orbits, and synchrotron radiation. The possibility of \emph{floating orbits} -- orbits which do not decay -- arises due to \emph{superradiance}, which allows a particle on a corotating orbit to stimulate the release of energy and angular momentum from a rotating black hole  \cite{Press:1972zz, Cardoso:2011xi, Kapadia:2013kf}. The possibility of synchrotron radiation arises from the high velocities on ISCO orbits, leading to the beaming of radiation in the direction of motion \cite{Misner:1972jf, Davis:1972dm}.

In 1960, DeWitt and Brehme \cite{DeWitt:1960fc} derived an expression for the self-force on a point electric charge (see Eq.~(1.33) in Ref.~\cite{Poisson:2011nh}) that consists of two parts: a local term which depends on the external force and the local Ricci tensor \cite{hobbs1968vierbein}, and a \emph{tail integral}, which encapsulates the effect of radiation emitted at earlier times that reaches the particle after interacting with the spacetime curvature. Thus, self-force in curved spacetime is non-local in time, since it depends on the past history of the motion of the particle, as well as its current state.

Calculating the tail integral in practice is a technical challenge (though see \cite{Wardell:2014kea}); fortunately, there are equivalent formulations available, as described in the review articles \cite{Poisson:2011nh} and \cite{Barack:2018yvs} (see also Ref.~\cite{Khusnutdinov:2020mkz}).  Prominent among these is 
the \emph{mode sum regularization} (MSR) method introduced by Barack and Ori \cite{Barack:1999wf}, which has been applied by numerous authors \cite{Barack:2007tm,Barack:2010tm,Akcay:2010dx,Shah:2010bi,Shah:2012gu,Akcay:2013wfa,Dolan:2012jg,Osburn:2014hoa,vandeMeent:2015lxa,vandeMeent:2017bcc} for efficient and accurate calculations of the self-force. Schematically, a regularized self-force $\mathcal{F}^{\text{reg}}_\mu$ is obtained by subtracting \emph{regularization parameters} $ \mathcal{F}^{[-1]}_\mu$, $\mathcal{F}^{[0]}_\mu$, etc., from the $\ell$ modes of a `bare' force:
\be
\mathcal{F}^{\text{reg}}_\mu = \sum_{\ell=0}^{\infty} \left( \mathcal{F}_\mu^{\text{bare}, \ell} - \mathcal{F}^{[-1]}_\mu (2 \ell+1) -  \mathcal{F}^{[0]}_\mu - \ldots  \right) . 
\ee
The regularization parameters are obtained from a local analysis of the symmetric-singular Detweiler-Whiting field \cite{Detweiler:2002mi}. Happily, regularization parameters for the EM field have already been calculated for the Schwarzschild black hole by Barack and Ori \cite{Barack:2002bt} and for the Kerr black hole by Heffernan, Wardell and Ottewill \cite{Heffernan:2012su,Heffernan:2012vj,Heffernan:2014uxa}, and we make use of these here.

The MSR method is suited to cases where the field equations allow for a complete decomposition into modes in such a way as to reduce the problem to the solution of ordinary differential equations. Fortunately, the field equations for an EM field on Kerr spacetime fall into this class, as shown by Teukolsky \cite{Press:1973zz,Teukolsky:1973ha,Teukolsky:1974yv}, and the Faraday tensor $F_{\mu \nu}$ can be fully reconstructed from Maxwell scalars of spin-weight $\pm 1$ that satisfy second-order ODEs \cite{chandrasekhar1976solution, Chandrasekhar:1985kt}. 

The article is organised as follows. Sec.~\ref{sec:formulation} describes the formulation of the calculation, covering the spacetime and its geodesic orbits (\ref{sec:spacetime}); Maxwell's equations in the Teukolsky formalism (\ref{sec:maxwell}); the distributional source terms due to the particle (\ref{sec:source-terms}); the mode solutions (\ref{sec:modes}) and the special cases of static modes and the monopole (\ref{sec:monopole}); the dissipative self-force and fluxes (\ref{sec:dissipative}); and the conservative self-force (\ref{sec:conservative}) calculated by projecting from spin-weighted spheroidal harmonics to spherical harmonics (\ref{sec:projection}) and by mode sum regularization (\ref{sec:regularization}). Sec.~\ref{sec:implementation} describes the implementation, addressing numerical issues (\ref{sec:numerics}) and the validation of the results (\ref{sec:validation}). Results are given in Sec.~\ref{sec:results} for the dissipative (\ref{sec:dis-results}) and conservative (\ref{sec:con-results}) aspects of the self-force. We conclude with a discussion in Sec.~\ref{sec:conclusion}. 

We employ units in which the physical constants $G$, $c$ and $4 \pi \epsilon_0$ are equal to unity. The spacetime signature is $\{-+++\}$. 

\section{Formulation\label{sec:formulation}}

 \subsection{Spacetime and geodesic orbits\label{sec:spacetime}}

\subsubsection{Spacetime}
The Kerr solution with mass $M$ and angular momentum $J = aM$ expressed in Boyer-Lindquist coordinates $\{t,r,\theta,\phi\}$ has the line element
\begin{align}
ds^2 \equiv g_{\mu \nu} dx^\mu dx^\nu =  -\frac{\Delta}{\Sigma} \left( d t-a\sin^2\theta d \phi \right)^2 + \frac{\Sigma}{\Delta} d r^2 + \Sigma\, d\theta^2 + 
 \frac{\sin^2\theta}{\Sigma}\left((r^2+a^2) d \phi - a dt\right)^2,\label{eq:linelement}
\end{align}
where $\Sigma\equiv r^2+a^2\cos^2\theta$ and $\Delta \equiv r^2-2Mr+a^2$.  When the condition $a^2 \leq M^2$ is satisfied, the Kerr solution corresponds to a black hole spacetime with two distinct horizons: an internal~(Cauchy) horizon at~$r_{-}=M-\sqrt{M^2-a^2}$ and an external~(event) horizon at~$r_{+}=M+\sqrt{M^2-a^2}$. The angular velocity of the event horizon is
\be
\Omega_h = \frac{a}{2Mr_{+}} . \label{eq:omegaH}
\ee

The inverse metric $g^{\mu \nu}$ can be written in terms of a null basis $\{l^\mu, n^\mu, m^\mu, \mbar^\mu\}$, where the overline denotes the complex conjugate, as
\begin{align}
g^{\mu \nu} &= -2 l^{(\mu} n^{\nu)} + 2 m^{(\mu} \mbar^{\nu)}  \\ 
 &= \frac{\Delta_r}{\Sigma} l_{+}^{(\mu} l_{-}^{\nu)} + \frac{1}{\Sigma} m_{+}^{(\mu} m_{-}^{\nu)} .
\end{align}
Here we employ the Kinnersley tetrad,
\begin{subequations}
\begin{align}
l^\mu &= l_+^\mu, & 
n^\mu &= - \frac{\Delta_r}{2\Sigma} l_-^\mu, &
m^\mu &= \frac{1}{\sqrt{2} (r + i a \cos \theta) } m_+^\mu,  
\end{align}
\label{eq:tetrad1}
\end{subequations}
written in terms of an non-normalised null basis
\begin{align}
l^\mu_\pm &\equiv \left[ \pm (r^2+a^2) / \Delta, 1, 0, \pm a / \Delta \right] , &
m^\mu_{\pm} &\equiv \left[ \pm i a \sin \theta, 0, 1, \pm i \csc \theta \right] = \overline{m}^\mu_{\mp}.
\label{eq:tetrad2}
\end{align}
The legs $l_\pm^\mu$ are aligned with the two principal null directions of the spacetime. The inner products of the tetrad $l_\pm^\mu$ and $m_\pm^\mu$ are
\begin{equation}
g_{\mu \nu} l_+^\mu l_-^\nu = \frac{2 \Sigma}{\Delta}, \quad \quad g_{\mu \nu} m_+^\mu m_-^\nu =  2 \Sigma,
\end{equation}
with all others zero. 

\subsubsection{Circular equatorial geodesic orbits\label{subsec:orbit}}
Let $x_p^\mu(\tau)$ denote the particle's worldline, with tangent vector $u^\mu \equiv \frac{d x_p^\mu}{d\tau}$ satisfying $g_{\mu \nu} u^\mu u^\nu = -1$. In the absence of forces $x_p^\mu(\tau)$ is a geodesic, satisfying $u^\nu \nabla_\nu u^\mu = 0$. Geodesic orbits on the Kerr spacetime are characterized by three constants of motion:  energy $E = -u_\mu \xi^\mu_{(t)}$,  azimuthal angular momentum $L = u_\mu \xi^\mu_{(\phi)}$ and Carter constant $Q = Q^{\mu \nu} u_{\mu} u_{\nu}$, where $\xi^\mu_{(t)} = (\partial_t)^\mu$ and $\xi^\mu_{(\phi)} = (\partial_\phi)^\mu$ are Killing vectors and $Q^{\mu \nu}$ is the Killing tensor. For a circular orbit in the equatorial plane at Boyer-Lindquist radius $r_0$, 
\beq
E = \frac{1 - 2 \nu^2 + \tilde{a} \nu^3}{\sqrt{1 - 3 \nu^2 + 2 \tilde{a} \nu^3}}, \quad \quad
L = r_0 \nu \frac{1 - 2 \tilde{a} \nu^3 + \tilde{a}^2 \nu^4}{\sqrt{1 - 3 \nu^2 + 2 \tilde{a} \nu^3}}, \quad \quad
Q = 0, 
\eeq
where $\tilde{a} = a/M$ and $\nu = \sqrt{M / r_0}$. Explicitly, the equatorial circular geodesic orbit has
$
x^\mu_p(\tau) = \left[ t(\tau), r_0, 0, \Omega t(\tau) \right] 
$
and
$
u^\mu = u^t \left[ 1, 0, 0, \Omega \right] , 
$
where
\begin{align}
\Omega &= \frac{\nu^3}{M(1 + \tilde{a} \nu^3)} ,  \label{eq:Omega} &
u^t &= \frac{1 + \tilde{a} \nu^3}{\sqrt{1 - 3 \nu^2 + 2 \tilde{a} \nu^3}} . 
\end{align}
We adopt the convention \cite{Warburton:2010eq} that $L$ and $\Omega$ are always positive and $a>0$ ($a < 0$) for prograde (retrograde) orbits.

The innermost stable circular orbit (ISCO) is at the radius
\beq
r_{isco}/M = 3 + Z_2 \mp \sqrt{(3-Z_1)(3+Z_1+2Z_2)}  \label{eq:isco}
\eeq
where $Z_1 = 1 + (1 - \tilde{a}^2)^{1/3} \left[ (1+\tilde{a})^{1/3} + (1-\tilde{a})^{1/3} \right]$ and $Z_2 = \sqrt{3 \tilde{a}^2 + Z_1^2}$ and the upper (lower) sign in Eq.~(\ref{eq:isco}) corresponds to prograde (retrograde) motion \cite{Bardeen:1972fi,Isoyama:2014mja}.

\subsection{Maxwell's equations and the Teukolsky formalism\label{sec:maxwell}}

The electromagnetic field equations in their standard covariant form are
\beq
\nabla_{\nu} F^{\mu \nu} = 4 \pi J^\mu, \quad \quad \nabla_{[\mu} F_{\nu \sigma]} = 0,   \label{fieldeq}
\eeq
where $F^{\mu \nu}$ is the Faraday tensor and $J^\mu$ is a vector field representing a four-current that is divergence-free ($\nabla_\mu J^\mu = 0$). It is convenient to introduce a complexified version of the Faraday tensor, $\cF^{\mu \nu} = F^{\mu \nu} - i \widetilde{F}^{\mu \nu}$, where $\widetilde{}$ denotes the Hodge dual, i.e., $\widetilde{F}^{\mu \nu} = \frac{1}{2} \varepsilon^{\mu \nu \sigma \gamma} F_{\sigma \gamma}$ . The complexified tensor is self-dual by virtue of the property $\widetilde{\cF}^{\mu \nu} = i\cF^{\mu \nu}$. The field equations (\ref{fieldeq}) then reduce to a single tensorial equation
\beq
\nabla_{\nu} \cF^{\mu \nu} = 4 \pi J^\mu. \label{fieldeq2}
\eeq

The six degrees of freedom of $F^{\mu \nu}$ are encapsulated in 3 complex Maxwell scalars, 
\beq
\phi_0 \equiv F_{\mu \nu} l^\mu m^\nu , \quad \phi_2 \equiv F_{\mu \nu} \mbar^\mu n^\nu , \quad \quad \phi_1 \equiv \frac{1}{2} F _{\mu \nu} \left( l^\mu n^\nu - m^\mu \mbar^\nu \right) ,
\label{eq:maxwelldef1}
\eeq
and the self-dual Faraday tensor is specified in terms of Maxwell scalars by
\beq
\cF^{\mu \nu} = 4 \left( \phi_0 \, \overline{m}^{[\mu} n^{\nu]} + \phi_2 \, l^{[\mu} m^{\nu]} + \phi_1 \, (n^{[\mu} l^{\nu]} - \overline{m}^{[\mu} m^{\nu]} ) \right) . \label{eq:selfdualF}
\eeq
For future reference, we introduce rescaled quantities:
\begin{subequations}
\begin{align}
\Phi_{+1} &\equiv \phi_0 = \frac{1}{\sqrt{2} \varrho} l_+^\mu m_+^\nu F_{\mu \nu}  & \Phi_{-1} &\equiv 2 \overline{\varrho}^2 \phi_2 = \frac{\Delta}{\sqrt{2} \varrho} l_-^\mu m_-^\nu F_{\mu \nu} , 
\end{align}
\end{subequations}
where $\varrho = r + i a \cos \theta$. 

Projecting (\ref{fieldeq2}) onto a null tetrad aligned with the principal null directions leads to four equations in Newman-Penrose form \cite{Teukolsky:1973ha}
\begin{subequations}
\begin{align}
(D - 2 \rho) \phi_1 - (\overline{\delta} + \pi - 2 \alpha) \phi_0 &= -2 \pi J_l , \\
(\delta - 2 \tau) \phi_1 - (\varDelta + \mu - 2 \gamma) \phi_0 &= -2 \pi J_m , \\
(D - \rho + 2\epsilon) \phi_2 - (\overline{\delta} + 2\pi) \phi_1 &= -2 \pi J_{\overline{m}} , \\
(\delta - \tau + 2 \beta) \phi_2 - (\varDelta + 2 \mu) \phi_1 &= -2 \pi J_n ,
\end{align}
\end{subequations}
where $D = l^\mu \partial_\mu$, $\varDelta = n^\mu \partial_\mu$, $\delta = m^\mu \partial_\mu$ are directional derivatives, and 
$J_l = l^\mu J_\mu$, $J_n = n^\mu J_\mu$, etc., are projections of the four-current, and $\alpha, \rho, \tau, \pi$ etc.~are the Newman-Penrose coefficients associated with the null tetrad. 

In 1973, Teukolsky \cite{Teukolsky:1973ha} showed that one can obtain a decoupled equation for $\phi_0$, and also for $\phi_2$, by exploiting a commutation relation between first-order operators. After inserting the Newman-Penrose quantities for the Kinnersley tetrad, viz.~$\kappa=\sigma=\nu=\lambda= 0$,
\begin{subequations}
\begin{align}
\rho &= -1/(r-ia\cos\theta) , & \beta &= -\rho^\ast \cot \theta / 2\sqrt{2}, & \pi &= i a \rho^2 \sin \theta / \sqrt{2}, & \alpha &= \pi - \beta^\ast , \\
\tau &= -ia\rho\rho^\ast \sin\theta / \sqrt{2}, & \mu &= \rho^2 \rho^\ast \Delta_r / 2, & \gamma &= \mu + \frac{1}{4}\rho \rho^\ast \Delta' , 
& \epsilon &= 0 ,
\end{align}
\label{eq:npcoeffs}
\end{subequations}
one arrives at a master equation, Eq.~(4.7) in Ref.~\cite{Teukolsky:1973ha}. This may be cast into the form \cite{Bini:2002jx}
\beq
\left[ (\nabla_{\mu} \pm \Gamma_\mu ) ( \nabla^{\mu} \pm \Gamma^\mu ) - 4 \psi_2  \right] \Phi_{\pm 1} = 4 \pi T_{\pm 1}, 
\label{eq:Bini}
\eeq
where $\nabla_\mu$ denotes the covariant derivative on the Kerr spacetime, and here the so-called ``connection vector'' \cite{Bini:2002jx} is 
\beq
\Gamma^\mu \equiv \frac{1}{\Sigma} \left[ \frac{M(r^2 - a^2)}{\Delta} - (r + i a \cos \theta) , r-M, 0, \frac{a(r-M)}{\Delta} + i \frac{\cos \theta}{\sin^2 \theta} \right] 
\eeq
and $\psi_2 = M / (r - i a \cos \theta)^3$ is the only non-vanishing Weyl scalar for the Kerr spacetime in the Kinnersley tetrad. 
The source terms in Eq.~(\ref{eq:Bini}) are 
\begin{align}
T_{+1} =  J_0 &\equiv  \left(\delta - \beta - \overline{\alpha} - 2 \tau + \overline{\pi} \right) J_l- \left(D - \epsilon + \overline{\epsilon} - 2 \rho - \overline{\rho} \right) J_m ,  \label{J0}  \\
\frac{1}{2 \overline{\varrho}^2} T_{-1} =  J_2 &\equiv  \left(\varDelta + \gamma - \overline{\gamma} + 2 \mu + \overline{\mu} \right) J_{\overline{m}} - \left( \overline{\delta} + \alpha + \overline{\beta} + 2 \pi - \overline{\tau} \right) J_n  .   \label{J2}
\end{align}
Remarkably, Eq.~(\ref{eq:Bini}) admits separable solutions. The solution can be constructed from a sum over modes, with each mode in the form
\begin{align}
\Phi_{\pm1} &= R_{\pm1}(r) S_{\pm1}(\theta) e^{-i \omega t + i m \phi} . \label{eq:separable}
\end{align}
In the vacuum case ($J^\mu = 0$), inserting Eq.~(\ref{eq:separable}) into Eq.~(\ref{eq:Bini}) leads to homogeneous Teukolsky equations in Chandrasekhar's form,
\begin{subequations}
\begin{align}
\left(\Delta \cD^\dagger \cD - 2 i \omega r  - \lambda\right) P_{-1} &= 0 ,    &
\left(\cL \cL_1^\dagger + 2 a \omega \cos \theta + \lambda \right) S_{-1} &= 0,   \label{eq:Teuk1}  \\
\left(\Delta \cD \cD^\dagger + 2 i \omega r - \lambda \right) P_{+1} &= 0 ,  &
\left(\cL^\dagger \cL_1 - 2 a \omega \cos \theta + \lambda \right) S_{+1} &= 0 ,  \label{eq:Teuk2} 
\end{align}
\label{eq:teukolsky-eqns}
\end{subequations}
where $P_{+1} = \Delta R_{+1}$, $P_{-1} = R_{-1}$ and $\lambda$ is the separation constant for $s=-1$ \cite{Chandrasekhar:1985kt}. Here we have made use of directional derivatives along $\{ l_+^\mu, l_-^\mu, m_+^\mu, m_-^\mu \}$, denoted by
 $\{ \cD, \cD^\dagger, \cL^\dagger, \cL \}$, where 
\begin{subequations}
\begin{align}
\cD \equiv l_+^\mu \partial_\mu &= \partial_r - \frac{i K}{\Delta},  & 
\cL^\dagger \equiv m_+^\mu \partial_\mu =  \partial_\theta - Q ,
 \\
\cD^\dagger \equiv l_-^\mu \partial_\mu &= \partial_r + \frac{i K}{\Delta} , 
& \cL \equiv m_-^\mu \partial_\mu = \partial_\theta + Q ,
\end{align}
\label{eq:DLoperators}
\end{subequations}
with $K \equiv \omega (r^2 + a^2) - a m$ and $Q \equiv m \csc \theta - a \omega \sin \theta$. We assume that these operators act only on quantities with harmonic time dependence $\chi \equiv e^{-i \omega t + i m \phi}$. Furthermore, let $\cL_n = \cL + n \cot\theta$ and $\cL^\dagger_n = \cL^\dagger + n \cot\theta$. 

For consistency these functions must also satisfy the Teukolsky-Starobinsky identities,
\begin{subequations}
\begin{align}
\Delta \mathcal{D} \mathcal{D} P_{-1} &= \mathcal{B} \, P_{+1} , &
\mathcal{L}^\dagger \mathcal{L}_1^\dagger S_{-1} &= \mathcal{B} \, S_{+1} ,  \label{TS1}  \\
\Delta \mathcal{D}^\dagger \mathcal{D}^\dagger P_{+1} &= \mathcal{B} \, P_{-1} , &
\mathcal{L} \mathcal{L}_1 S_{+1} &= \mathcal{B} \, S_{-1} ,  \label{TS2} 
\end{align}
\label{TS}
\end{subequations}
where 
$
\mathcal{B} \equiv \sqrt{\lambda^2 + 4 a m \omega - 4 a^2 \omega^2} \label{eq:Bteuk}
$.

The modes of the Maxwell scalar of zero spin-weight, $\phi_1$, can be constructed by applying differential operators to the modes of $\phi_0$ and $\phi_2$ \cite{chandrasekhar1976solution}. From Chap.~8 in Chandrasekhar \cite{Chandrasekhar:1998}, 
\begin{subequations}
\begin{align}
\phi^{\ell m}_1 &= \frac{\chi}{\sqrt{2} (r-i a\cos\theta)^2} \left[ g_{+1}(r) \cL_1 S_{+1}(\theta) - i a f_{-1}(\theta) \cD P_{-1}(r) \right]   \label{eq:phi1chandra} \\
 &=  -\frac{\chi}{\sqrt{2} (r-i a\cos\theta)^2} \left[ g_{-1}(r) \cL^\dagger_1 S_{-1}(\theta) - i a f_{+1}(\theta) \cD^\dagger P_{+1}(r)  \right] 
\end{align}
\end{subequations}
where
\begin{subequations}
\begin{align}
\cB \, g_{+1}(r) &= \left( r \cD - 1 \right) P_{-1} , \\
\cB \, g_{-1}(r) &= \left( r \cD^\dagger - 1 \right) P_{+1} , \\
\cB \, f_{+1}(\theta) &= \left( \cos \theta \cL_1^\dagger + \sin \theta \right) S_{-1} , \\
\cB \, f_{-1}(\theta) &= \left( \cos \theta \cL_1 + \sin \theta \right) S_{+1} .
\end{align}
\end{subequations}

\subsection{Source terms\label{sec:source-terms}}

For a point-like charge $q$ on a geodesic orbit, the four-current is
\begin{align}
J^\mu &= q \int u^\mu(\tau) \delta^4 \left( x^\mu - x_p^\mu(\tau) \right) (-g(x))^{-1/2} d\tau  , \\
 &= \frac{q \hat{U}^\mu}{r_0^2} \delta(r - r_0) \delta(\theta - \pi/2) 
 \delta(\phi - \Omega t) .
\end{align}
On the second line we have inserted the expressions in Sec.~(\ref{subsec:orbit}) to specialise to a circular geodesic orbit in the equatorial plane ($\theta=\pi/2$). Here $\hat{U}^\mu \equiv u^\mu / u^t = [1,0,0,\Omega]$, with $\Omega$ defined in Eq.~(\ref{eq:Omega}); projecting onto the Kinnersley tetrad yields
\begin{align}
\hat{U}^\mu l_\mu &= -(1-a\Omega) = \hat{U}^\mu n_\mu \frac{2 r_0^2}{\Delta_0} , &
\hat{U}^\mu m_\mu &= \frac{i}{\sqrt{2} \, r_0} \left( (r_0^2+a^2) \Omega - a \right) .
\end{align}
The first task is to compute the source terms $J_0$ and $J_2$ in Eqs.~(\ref{J0}) and (\ref{J2}). Here we must handle the distributional terms with some care, noting that whereas $f(x) \delta(x - x_0) = f(x_0) \delta (x - x_0)$, on the other hand
\beq
f(x) \delta'(x - x_0) = f(x_0) \delta'(x-x_0) - f'(x_0) \delta(x - x_0) , \label{eq:deltaidentity}
\eeq
where $f(x)$ is any differentiable function and $x_0$ is a constant. Using
\beq
\delta(\phi - \Omega t) = \frac{1}{2 \pi} \sum_{m=-\infty}^{\infty} \chi_m , \quad \quad \chi_m \equiv e^{i m (\phi - \Omega t)} ,
\eeq
and evaluating on the equatorial plane at $r=r_0$ after employing (\ref{eq:deltaidentity}) leads to
\begin{align}
\Sigma J_0 &=\frac{-q}{2 \pi \sqrt{2} r_0} \sum_{m}  \left[ (1-a\Omega) \left( \partial_\theta - m(1-a\Omega) + \frac{i a}{r_0} \right) + \right.  \\ 
& \quad \quad \quad \quad  \quad \quad \left. + i ((r_0^2+a^2)\Omega - a) \left( \partial_r - \frac{i m ((r_0^2+a^2)\Omega - a)}{\Delta_0} + \frac{1}{r_0} \right) \right] \chi_m \delta(r-r_0) \delta(\theta - \tfrac{\pi}{2}) . \nn 
\end{align}
At this point we employ the orthonormality of the spin-weighted spheroidal harmonics,
\beq
\int S_{\pm1}^{\ell m}(\theta) S_{\pm1}^{\ell'm}(\theta) d(\cos \theta) = \frac{1}{2 \pi} \delta_{\ell \ell'} ,  \label{eq:Snorm}
\eeq
to establish that
\begin{align}
\delta(\theta - \tfrac{\pi}{2}) &= 2 \pi \sum_{\ell=1}^\infty S_{\pm1}^{\ell m}(\tfrac{\pi}{2}) S_{\pm1}^{\ell m}(\theta) , \\
\delta'(\theta - \tfrac{\pi}{2}) &= 2 \pi \sum_{\ell=1}^\infty -S^{\ell m \, \prime}_{\pm1} (\tfrac{\pi}{2}) S_{\pm1}^{\ell m}(\theta) .
\end{align}
Hence
\begin{align}
\Sigma J_0 &= \frac{-q}{\sqrt{2} r_0} \sum_{\ell m} S_{+1}^{\ell m}(\theta) \chi_m \left\{ i ((r_0^2 + a^2) \Omega - a) S_{+1}^{\ell m}(\tfrac{\pi}{2}) \delta'(r-r_0) + \phantom{\frac{blah}{blah}} \right. \nn \\
& \quad \quad \quad \quad \quad \quad \quad \quad \quad - (1 - a \Omega) S_{+1}^{\ell m \, \prime}(\tfrac{\pi}{2}) \delta(r-r_0) + \nn \\
& \left. \quad \quad \quad \quad \quad \quad \quad \quad \quad + \left[ i r_0 \Omega  + m \left(\frac{ ((r_0^2+a^2)\Omega - a)^2}{\Delta_0} - (1-a\Omega)^2 \right) \right] S_{+1}^{\ell m}(\tfrac{\pi}{2})  \delta(r-r_0) \right\} . \label{source:J0}
\end{align}

From the form of (\ref{source:J0}), we see that the master equation Eq.~(\ref{eq:Bini}) admits a separable solution
\beq
\Phi_{\pm 1} = \sum_{\ell = 1}^\infty \sum_{m = -\ell}^{\ell} R_{\pm1}^{\ell m} S_{\pm 1}^{\ell m} \chi_m
\eeq
where 
\begin{subequations}
\begin{align}
\left(\Delta \cD \cD^\dagger + 2 i m \Omega  r - \lambda \right) P^{\ell m}_{+1} &= \mathcal{S} \left(+ i B S_+ \delta^\prime(r - r_0) + \left\{ (m A^{(r)} + i A^{(i)}) S_+ + C S_+^\prime \right\} \delta(r-r_0) \right) ,  \label{eq:TeukRp}  \\
\left(\Delta \cD^\dagger \cD - 2 i m \Omega r  - \lambda \right) P^{\ell m}_{-1} &=  \mathcal{S} \left(- i B S_- \delta^\prime(r - r_0) + \left\{ (m A^{(r)} - i A^{(i)}) S_- - C S_-^\prime \right\} \delta(r-r_0) \right) ,   \label{eq:TeukRm} 
\end{align}
\label{eq:Teuksrc}
\end{subequations}
where $P^{\ell m}_{+1} = \Delta R^{\ell m}_{+1}$ and $P^{\ell m}_{-1} = R^{\ell m}_{-1}$, and $S_{\pm 1} = S^{\ell m}_{\pm1}(\tfrac{\pi}{2})$ and $S^\prime_{\pm 1} = S^{\ell m \, \prime}_{\pm1}(\tfrac{\pi}{2})$, and
\begin{subequations}
\begin{align}
\mathcal{S} &= \bl{-} \frac{4 \pi q}{\sqrt{2} r_0} , \\
B &= \Delta_0 ((r_0^2 + a^2) \Omega - a) , \\
A^{(r)} &= r_0 \left( r_0 \left((r_0^2+a^2) \Omega^2 - 1 \right) + 2M(1-a\Omega)^2 \right), \\
A^{(i)} &= a^2 (2M - r_0) \Omega + 2 a (r_0 - M) - r_0^3 \Omega , \\
C &= - \Delta_0 (1 - a \Omega) .
\end{align}
\label{eq:ABC}
\end{subequations}

\subsection{Mode solutions\label{sec:modes}}
The source terms in Eqs.~(\ref{eq:Teuksrc}) are distributions with support at $r=r_0$ only. Hence solutions to the inhomogeneous equations may be constructed from solutions to the homogeneous equations in the standard manner. Let 
$\Pin_{\pm 1}$ and $\Pup_{\pm 1}$ be a pair of solutions to Eq.~(\ref{eq:teukolsky-eqns}) that satisfy the physical boundary conditions, that is, let $\Pin_{\pm 1}$ be ingoing at the future horizon, and let $\Pup_{\pm 1}$ be outgoing at future infinity. The inhomogeneous solution takes the form
\beq
P^{\ell m}_{\pm1}(r) = \alpha_{\pm1}^{\infty} \Pup_{\pm1}(r) \Theta(r-r_0) + \alpha_{\pm1}^h \Pin_{\pm1}(r) \Theta(r_0 - r) , \label{eq:inhomog}
\eeq
where $\Theta(\cdot)$ is the Heaviside step function, and $\alpha_{\pm1}^{\infty}$ and $\alpha_{\pm1}^h$ are complex coefficients to be determined. Inserting (\ref{eq:inhomog}) into (\ref{eq:Teuksrc}) yields the matrix equation
\beq
\begin{pmatrix} \alpha^{\ell m, \infty}_{\pm1} \\ \alpha^{\ell m, h}_\pm \end{pmatrix} =
\frac{1}{\Delta_0 W_{\pm}} \left. \begin{pmatrix} -(\Pin_{\pm1})^\prime & \Pin_{\pm1} \\ -(\Pup_{\pm1})^\prime & \Pup_{\pm1} \end{pmatrix} \right|_{r=r_0}
\begin{pmatrix} \mathfrak{B}_{\pm} \\ \mathfrak{A}_{\pm} \end{pmatrix} .
\eeq
where 
\begin{subequations}
\begin{align}
W_{\pm} &= \Pin_{\pm1} \frac{d \Pup_{\pm1}}{dr} - \Pup_{\pm1} \frac{d \Pin_{\pm1}}{dr} , \\
\mathfrak{B}_{\pm} &= \pm i \mathcal{S} B S_{\pm}  , \\
\mathfrak{A}_{\pm} &= \mathcal{S} \left\{ (m A^{(r)} \pm i \widetilde{\mathcal{A}}^{(i)}) S_\pm \pm C S_\pm^\prime \right\} .
\end{align}
\label{eq:radialsourceterms}
\end{subequations}
Here $\mathcal{S}$, $B$, $A^{(r)}$ and $C$ are defined in Eq.~(\ref{eq:ABC}), and $\widetilde{\mathcal{A}}^{(i)} = r_0 \Delta_0 \Omega$. 

\subsection{Static modes and the monopole\label{sec:monopole}}

\subsubsection{m=0 homogeneous modes}
The $m=0$ modes are static ($\omega = 0$). In this case we employ the homogeneous modes
\begin{align}
\Pinstatic_{\pm 1} &= \Delta \partial_r P_{\ell}(z) , &
\Pupstatic_{\pm 1} &= \Delta \partial_r Q_{\ell}(z) ,
\end{align}
where $P_{\ell}(\cdot)$ and $Q_{\ell}(\cdot)$ are Legendre functions with the branch cut on the real axis from $-\infty$ to $+1$, and $z \equiv \Delta_{,r} / (r_+ - r_-)$. The Wronskian is
\beq
W_{\pm} \equiv \Pinstatic_{\pm1} \frac{d \Pupstatic_s}{dr} - \Pupstatic_s \frac{d \Pinstatic_s}{dr} = \frac{1}{2} (r_+ - r_-) \ell (\ell + 1) . 
\eeq
The angular functions are
\beq
S^{\ell 0}_{\pm 1}(\theta) = \mp \sqrt{\frac{2 \ell + 1}{4 \pi \ell (\ell + 1)}} \frac{d}{d \theta} P_{l}(\cos \theta) , \label{eq:Sl0}
\eeq 
such that the normalisation condition (\ref{eq:Snorm}) holds. 

\subsubsection{Monopole mode}
To complete the solution, we must now add `by hand' a non-radiative monopole mode which is responsible for the $q/r$ part of the electric field far from the black hole. 

The homogeneous vector potential 
\beq
A_{(0)}^{\mu} \equiv \frac{q r}{2 \Sigma} \left( l_{+}^\mu - l_{-}^\mu \right), 
\eeq
in Lorenz gauge ($\nabla_\mu A_{(0)}^\mu =0$) generates a homogeneous Faraday tensor $F_{(0)}^{\mu \nu} = \nabla^{\mu} A^\nu - \nabla^{\nu} A^\mu$ that satisfies the vacuum equation $\nabla_{\nu} F_{(0)}^{\mu \nu} = 0$. It has the key properties that
\beq
F_{(0)}^{tr} = \frac{q}{r^2} + O(r^{-3}) \, , \quad \quad
\eeq
in the far-field and
\beq
\frac{1}{2} \int F_{(0)}^{\mu \nu} dS_{\mu \nu} = 4 \pi q ,
\eeq
where the two-surface integral is taken over any `sphere' of constant Boyer-Lindquist coordinate $r$, or any closed surface enclosing the horizon. It is quick to verify that the Maxwell scalars $\phi_0$ and $\phi_2$ (but not $\phi_1$) associated with the homogeneous solution are zero. 

The inhomogeneous monopole mode, 
\beq
F^{\mu \nu}_{\text{mono}} = \Theta(r - r_0) F^{\mu \nu}_{(0)}  ,
\eeq
does \emph{not} satisfy the vacuum equation; instead, $\nabla_{\nu} \mathbb{F}_{\text{mono}}^{\mu \nu} = 4 \pi \mathbb{J}_{\text{mono}}^\mu$ where $\mathbb{F}_{\text{mono}}^{\mu \nu} \equiv F_{\text{mono}}^{\mu \nu} - i \widetilde{F}_{\text{mono}}^{\mu \nu}$ and it is straightforward to show that
\beq
\mathbb{J}_{\text{mono}}^\mu = \frac{q}{4 \pi} \frac{\Delta}{2 \Sigma} \frac{1}{(r-i a \cos \theta)^2} \left( l_{+}^\mu  - l_{-}^\mu \right) \delta(r - r_0) .
\eeq
Note that $\mathbb{J}_{\text{mono}}^\mu$ associated with the step in the monopole mode is not restricted to the particle worldline, but instead has support on the sphere at $r=r_0$. Although $\mathbb{J}_{\text{mono}}^\mu$ itself is not zero, a short calculation shows that there are no additional source terms for the Teukolsky equation (\ref{eq:Bini}), that is, $J_0^{\text{mono}} = J_2^{\text{mono}} = 0$. In other words, the inhomogeneous monopole is associated with a step in $\phi_1$, the Maxwell scalar of spin-weight zero, only.

The inhomogenous monopole mode makes a contribution to the radial component of the self-force of
\be
\mathcal{F}_r^{\text{mono}} = q^2 u^t \frac{(r_0^2 - a^2 \cos^2 \theta)(1 - a \Omega \sin^2 \theta)}{\Sigma^2} \equiv \sum_{\ell =0}^\infty \mathcal{F}_r^{\ell, \text{mono}} Y_0^{\ell 0}(\theta) .
\label{eq:monopole-force}
\ee
Evaluating at $\theta=\pi/2$ yields $\mathcal{F}_r^{\text{mono}} = q^2 u^t (1 - a \Omega) / r_0^2$.

 \subsection{Dissipative force and fluxes\label{sec:dissipative}}
 
 	\subsubsection{Dissipative component of the self-force}
 	
The dissipative components of the self-force are the $t$ and $\phi$ components of $\mathcal{F}_{\mu} \equiv q F_{\mu\nu}u^{\nu}$.
From the symmetry of the Faraday tensor, it is straightforward to see that $\mathcal{F}_{t} = q F_{t \phi} \Omega u^t = - \mathcal{F}_{\phi}\Omega$ and in the following we will focus on the $t$ component of the self-force.
The $(t\phi)$ component of the Faraday tensor can be expressed in terms of the Maxwell scalars as:
\begin{equation}
F_{t\phi} = \sqrt{2} \, \text{Re}\left[ i\sin \theta \, (r - i a \cos \theta)\phi_2 + \frac{i\Delta \sin \theta }{2\Sigma}(r + ia\cos \theta) \phi_0\right].
\end{equation}
Evaluating the force on the particle's worldline, i.e. at $r=r_0$ and $\theta = \pi/2$, yields
\begin{eqnarray}
\mathcal{F}_t &=& \sqrt{2} \, q u^t\Omega\, \text{Re}\left[ ir_0\phi_2 + \frac{i\Delta_0}{2r_0}\phi_0\right] \\
&=&  \frac{q\Omega u^t}{\sqrt{2}r_0} \sum_{\ell m} \text{Re} \left[iP^{\ell m}_{-1}(r_0)\Sminus^{\ell m}(\tfrac{\pi}{2}) + iP^{\ell m}_{+1}(r_0)\Splus^{\ell m}(\tfrac{\pi}{2})\right] \\
&=& \frac{q\Omega u^t}{\sqrt{2}r_0}  \sum_{\ell m} \text{Re} \left[i\left((-1)^{\ell +m}P^{\ell m}_{-1}(r_0) + P^{\ell m}_{+1}(r_0)\right)\Splus^{\ell m}(\tfrac{\pi}{2}) \right],\label{Ft_numerics}
\end{eqnarray}
where we have used the fact that $\Sminus^{\ell m}(\frac{\pi}{2}) = (-1)^{l+m}\Splus^{\ell m}(\frac{\pi}{2})$. 	
 	
 	\subsubsection{Energy flux}
 	
 	For an electromagnetic field given by a Faraday tensor $F_{\mu\nu}$ with energy-momentum $T^{\mu\nu} = F^{\mu\alpha}F^{\nu\beta}g_{\alpha\beta} - \frac{1}{4}F^{\alpha\beta}F_{\alpha\beta}g^{\mu\nu}$,  and a Killing vector $K^{\mu}$, one can construct a current:
\begin{equation}
Y^{\mu} = T^{\mu\nu}K_{\nu}.
\end{equation}
In vacuum, this current is divergence-free but in the presence of a source, which is the case of interest here, the current satisfies the following continuity equation:
\begin{equation}
\nabla_\mu Y^{\mu} = F^{\mu\nu}K_\mu J_\nu = \frac{\mathcal{F}_\mu K^{\mu}}{r_0^2 u^t} \delta(r-r_0)\delta(\theta - \frac{\pi}{2}) \delta(\phi - \Omega t).
\end{equation}
Using Gauss' theorem
\begin{equation}
\int_{V} \nabla_\mu Y^{\mu} \sqrt{-g}d^4x = \int_{\partial V}{Y^{\mu} d\Sigma_\mu}
\end{equation}
where $V$ is a space-time volume with boundary $\partial V$ that spans from the horizon to infinity, we can relate the force at the particle to the fluxes through the boundary.
Since the system is stationary, only the fluxes at infinity and through the horizon contribute to the total flux (see Appendix~\ref{A:flux}):
\begin{equation}\label{conservation_law}
\frac{\mathcal{F}_a K^{a}}{u^t} = \Phi^{K}_{\infty} + \Phi^{K}_{h},
\end{equation}
where the superscript $K$ correspond to the choice of Killing vector. As mentioned the link between the $t$ and $\phi$ component of the force is trivial and we focus on the time component of the force which correspond to the choice $K^{a} = [1,0,0,0]$. In the following we will drop the superscript $K$ and keep in mind that we are considering the energy flux.
In Appendix~\ref{A:flux} we derive the expression for the energy flux at infinity and through the horizon in terms of the $\alpha$ coefficients defined in Eq.~(\ref{eq:inhomog}).
Explicitly, the energy flux at infinity is
\begin{equation}
\Phi_\infty = \frac{1}{8 \pi} \sum_{\ell m} | \alpha_{-1}^{\ell m, \infty} |^2,
\end{equation}
and through the horizon,
\begin{equation}
\Phi_h = \frac{1}{8 \pi} \sum_{\ell m} {\frac{\omega}{2 M r_{+} \widetilde{\omega}} | \alpha_{+1}^{\ell m, h} |^2},
\end{equation}
with $\widetilde{\omega} = \omega - m\Omega_h$ and $\Omega_h$ as defined in Eq.~(\ref{eq:omegaH}).

 \subsection{Conservative force and regularisation\label{sec:conservative}}
 	\subsubsection{Conservative component of the self-force}
 	We compute here the conservative component of the self-force, i.e. $\mathcal{F}_r$, in terms of the Maxwell scalars. From the definition of the force, we have:
\begin{equation}
\mathcal{F}_r = q F_{r \mu}u^\mu = q u^t\left(F_{rt} + F_{r\phi}\Omega\right).
\end{equation}
Using the expression of the Faraday tensor in terms of the Maxwell scalars,
\begin{equation}
F_{\mu\nu} = 2\left[ \phi_2 l_{[\mu}m_{\nu]} + \phi_0 \overline{m}_{[\mu}n_{\nu]} + \phi_1\left( n_{[\mu}l_{\nu]} + m_{[\mu}\overline{m}_{\nu]} \right)\right] + c.c.,
\end{equation}
we get that
\begin{equation}
\frac{\mathcal{F}_r}{q u^t} = \sqrt{2} \left( (a^2 + r^2)\Omega - a\right) \sin \theta \left[ - \frac{i \phi_0}{4(r - i a\cos\theta)} + \frac{i \phi_2 (r - i a \cos \theta)}{2\Delta}\right] + (1-a\Omega \sin^2 \theta)  \phi_1 + c.c.
\end{equation}
Inserting the mode decompositions (\ref{eq:separable}) and (\ref{eq:phi1chandra}) and evaluating at $\phi = \Omega t$ yields
\begin{align}
\mathcal{F}_r &= q u^t\sum_{lm} \frac{\sqrt{2}\left( (r^2 + a^2)\Omega - a\right)}{4 \Delta (r -ia\cos\theta)}\sin(\theta)\left[ -iP_{+1}^{lm} S_{+1}^{lm} + i P_{-1}^{lm} S_{-1}^{lm} \right] \nn \\ 
 & \quad \quad \quad \quad \quad + (1-a\Omega \sin^2\theta)\frac{g_{+1}^{lm}\mathcal{L}_1 S_{+1}^{lm} - iaf_{-1}^{lm} \mathcal{D}P_{-1}^{lm}}{\sqrt{2}(r-ia\cos\theta)^2} + c.c.
\end{align}

 \subsubsection{Projection onto scalar harmonics\label{sec:projection}}
Before the mode sum regularization procedure can be applied, it is necessary to project the spin-weighted spheroidal harmonics onto the scalar spherical harmonics. Using the results of Appendix~\ref{A:projection},
\begin{eqnarray}
\mathcal{F}_r = & & q u^t \sum_{l,m,\hat{l},\tilde{l}} \left\lbrace \frac{\sqrt{2}\left( (r^2 + a^2)\Omega - a\right)}{4 \Delta (r -ia\cos\theta)} \left[ - iP_{+1}^{lm} \mathcal{C}^{+1}_{lm\hat{l}\tilde{l}} + i P_{-1}^{lm} \mathcal{C}^{-1}_{lm\hat{l}\tilde{l}} \right] \right. \\
&& \quad \quad \quad \quad \quad + \left. (1-a\Omega \sin^2\theta)\frac{\mathcal{B} g_{+1}^{lm} \mathcal{C}^{\mathcal{L}}_{lm\hat{l}\tilde{l}} - ia\left[  \mathcal{C}^{\mathcal{L}}_{lm\hat{l}\tilde{l}} \cos\theta + \mathcal{C}^{+1}_{lm\hat{l}\tilde{l}} \right] \mathcal{D}P_{-1}^{lm}}{\sqrt{2}\mathcal{B}(r-ia\cos\theta)^2} \right\rbrace Y_0^{\tilde{l}m} + c.c.  \label{eq:Fr-projected}
\end{eqnarray}
with
\begin{subequations}
\begin{eqnarray}
\mathcal{C}^{+1}_{lm\hat{l}\tilde{l}} &=& \left( b_{+1}^m \right)^{l}_{\hat{l}} \left( A_{+1}^{m} \right)_{\tilde{l}}^{\hat{l}} \\
\mathcal{C}^{-1}_{lm\hat{l}\tilde{l}} &=& \left( b_{-1}^m \right)^{l}_{\hat{l}} \left( A_{-1}^{m} \right)_{\tilde{l}}^{\hat{l}} \\
\mathcal{C}^{\mathcal{L}}_{lm\hat{l}\tilde{l}} &=& \left( b_{+1}^m \right)^{l}_{\hat{l}} \left[ \sqrt{\hat{l}(\hat{l} +1)} \, \delta^{\hat{l}}_{\tilde{l}} - a m \Omega  (A_{+1}^{m})_{\tilde{l}}^{\hat{l}} \right] . 
\end{eqnarray}
\end{subequations}

Expanding (\ref{eq:Fr-projected}) in $z = \cos \theta$, we have
\begin{equation}
\mathcal{F}_r = q u^t \sum_{\ell m} \left[ {}_{0}\mathcal{F}_r^{\ell m} + {}_{1}f_r^{\ell m} z + {}_{2}f_r^{\ell m} z^2 + o(z^3) \right] Y_0^{\ell m} ,
\end{equation}
with
\begin{eqnarray}
{}_{0}\mathcal{F}_r^{\ell m} = \sum_{l \hat{l}}&& \frac{\left( (r_0^2 + a^2)\Omega - a\right)}{\sqrt{2}\Delta_0 r_0}\text{Im}\left[ P_{+1}^{lm}  \mathcal{C}^{+1}_{lm\hat{l}\ell}  - P_{-1}^{lm}  \mathcal{C}^{-1}_{lm\hat{l}\ell} \right] \nonumber \\
&+& \frac{\sqrt{2}(1-a\Omega)}{r_0^2}
\text{Re}\left[ g_{+1}^{lm} \mathcal{C}^{\mathcal{L}}_{lm\hat{l}\ell} - \frac{ia \mathcal{C}^{+1}_{lm\hat{l}\ell} \mathcal{D}P_{-1}^{lm}}{\mathcal{B}}\right],
\end{eqnarray}
and
\begin{eqnarray}
{}_{1}f_r^{\ell m} = a \, \sum_{l \hat{l}}&&\frac{( (r_0^2 + a^2)\Omega - a)}{\sqrt{2}\Delta_0 r_0^2} \, \text{Re}\left[ P_{+1}^{lm} \mathcal{C}^{+1}_{lm\hat{l}\ell}  - P_{-1}^{lm} \mathcal{C}^{-1}_{lm\hat{l}\ell} \right] 
\nonumber \\
 &-& \frac{2\sqrt{2}(1-a\Omega)}{r_0^3}\text{Im}\left[g_{+1}^{lm} \mathcal{C}^{\mathcal{L}}_{lm\hat{l}\ell} - \frac{ia}{\mathcal{B}} \mathcal{C}^{+1}_{lm\hat{l}\ell}  \mathcal{D}P_{-1}^{lm}\right] + 
  \frac{2 (1-a\Omega)}{\sqrt{2} r_0^2} \text{Im}\left[  \frac{\mathcal{C}^{\mathcal{L}}_{lm\hat{l}\ell} \mathcal{D}P_{-1}^{lm} }{\mathcal{B}}\right] , \\
{}_{2}f_r^{\ell m} = a^2  \, \sum_{l \hat{l}}&& \frac{- \left( (r_0^2 + a^2)\Omega - a\right)}{\sqrt{2} \Delta_0 r_0^3}\text{Im}\left[ P_{+1}^{lm}  \mathcal{C}^{+1}_{lm\hat{l}\ell}  - P_{-1}^{lm}  \mathcal{C}^{-1}_{lm\hat{l}\ell} \right] \nonumber \\
&& - \frac{3 \sqrt{2} (1-a\Omega)}{r_0^4}
\text{Re}\left[ g_{+1}^{lm} \mathcal{C}^{\mathcal{L}}_{lm\hat{l}\ell} - \frac{ia \mathcal{C}^{+1}_{lm\hat{l}\ell} \mathcal{D}P_{-1}^{lm}}{\mathcal{B}}\right]  + \frac{2 \sqrt{2} (1 - a \Omega)}{r_0^3} \text{Re} \left[  \frac{\mathcal{C}^{\mathcal{L}}_{lm\hat{l}\ell} \mathcal{D}P_{-1}^{lm} }{\mathcal{B}}\right] \nn \\
+ a \Omega \, \sum_{l \hat{l}}&& \frac{\sqrt{2}}{r_0^2}
\text{Re}\left[ g_{+1}^{lm} \mathcal{C}^{\mathcal{L}}_{lm\hat{l}\ell} - \frac{ia \mathcal{C}^{+1}_{lm\hat{l}\ell} \mathcal{D}P_{-1}^{lm}}{\mathcal{B}}\right],
\end{eqnarray}
Finally, expanding $z Y_0^{\ell m}$ and $z^2 Y_0^{\ell m}$ using
\begin{subequations}
\begin{align}
\cos \theta \, Y_0^{\ell m} &= \sum_{\ell_1} {}_{1}B^{\ell m}_{\ell_1} Y_0^{\ell_1 m}, \\
\cos^2 \theta \, Y_0^{\ell m} &= \sum_{\ell_2} {}_{2}B^{\ell m}_{\ell_2} Y_0^{\ell_2 m}, 
\end{align}
\end{subequations}
where
\begin{eqnarray}
{}_{1}B^{\ell m}_{\ell_1} &=& (-1)^m \sqrt{(2\ell + 1)(2\ell_1 + 1)} 
\begin{pmatrix}
1 & \ell & \ell_1 \\
0 & 0 & 0 
\end{pmatrix}
\begin{pmatrix}
1 & \ell & \ell_1 \\
0 & m & -m 
\end{pmatrix} \\
 {}_{2}B^{\ell m}_{\ell_2} &=& (-1)^m \frac{2 \sqrt{(2\ell + 1)(2\ell_2 + 1)}}{3}
\begin{pmatrix}
2 & \ell & \ell_2 \\
0 & 0 & 0 
\end{pmatrix}
\begin{pmatrix}
2 & \ell & \ell_2 \\
0 & m & -m 
\end{pmatrix}
+ \frac{1}{3}\delta_{\ell,\ell_2}
\end{eqnarray}
leads to
\begin{subequations}
\begin{align}
\mathcal{F}_r &= q u^t \sum_{\ell m} \left[ {}_{0}\mathcal{F}_r^{\ell m} + {}_{1}\mathcal{F}_r^{\ell m} + {}_{2}\mathcal{F}_r^{\ell m} + o(z^3) \right] Y^{\ell m}_0 , \\
 &= \sum_{\ell = 0}^\infty \mathcal{F}_r^{\ell} 
\end{align}
 \label{eq:Frl-kerr}
\end{subequations}
with
\begin{subequations}
\begin{align}
{}_{1}\mathcal{F}_r^{\ell m} &= \sum_{\ell_1} {}_{1}f_r^{\ell_1 m} \ {}_{1}B^{\ell_1 m}_{\ell} \\
{}_{2}\mathcal{F}_r^{\ell m} &= \sum_{\ell_2} {}_{1}f_r^{\ell_2 m} \ {}_{2}B^{\ell_2 m}_{\ell}. 
\end{align}
\end{subequations}

 \subsubsection{Mode sum regularization\label{sec:regularization}}

The regularization procedure is based on the subtraction of an appropriate singular component from the retarded field, in order to leave a finite regular field that is solely responsible for the self-force. The subtracted component must have the same singular structure as the retarded field in the vicinity of the particle, and must be sufficiently symmetric as to not contribute to the self-force (or at least, not in such a way that cannot be easily corrected for). Detweiler and Whiting identified an appropriate choice of the singular ($S$) field, based on a Green's function decomposition \cite{Detweiler:2002mi}. Subtracting this singular field is equivalent to regularizing at the level of the $\ell$-mode sum 
\cite{Barack:1999wf, Barack:2007tm,Barack:2010tm,Akcay:2010dx,Shah:2010bi,Poisson:2011nh,Shah:2012gu,Akcay:2013wfa,Dolan:2012jg,Osburn:2014hoa,vandeMeent:2015lxa,vandeMeent:2017bcc,Barack:2018yvs}. 

In the electromagnetic case, Heffernan \emph{et al.}~\cite{Heffernan:2012su,Heffernan:2012vj,Heffernan:2014uxa} (see also Haas \cite{Haas:2011np,Nolan-thesis}) showed that subtracting the $S$ field leads to a regularized force $\mathcal{F}^{\text{reg}}_\mu$ with a radial component in the form
\begin{align}
\mathcal{F}^{\text{reg}}_r &= \sum_{\ell = 0}^\infty \mathcal{F}_r^{\text{reg}[n]\ell}, \quad \quad \quad \mathcal{F}_r^{\text{reg}[n]\ell} \equiv \mathcal{F}_r^{\ell} - \mathcal{F}^{[n] \ell}_r  ,  \label{eq:Fr-reg}
\end{align}
where $[n]$ denotes the order of the local expansion of the $S$ field, and
\begin{equation}
\mathcal{F}^{[n] \ell}_r = (2 \ell + 1) \mathcal{F}_r^{[-1]} + \mathcal{F}_r^{[0]} + \frac{ \mathcal{F}_r^{[2]} }{(2\ell - 1)(2 \ell + 3)} + \ldots + \mathcal{F}_r^{[n]} \mathcal{G}_{[n]}(\ell). \label{eq:regparams}
\end{equation}
Here $n \ge 0$ is an even integer denoting the order, and $\mathcal{G}_{[n]}(\ell) \equiv 1/(2 \ell + 1 - n)(2 \ell +3 - n) \ldots (2 \ell + 1 + n)$ is defined for $n>0$ such that $\sum_{\ell =0}^\infty \mathcal{G}_{[n]}(\ell) = 0$. 
Explicit expressions for the mode sum regularization parameters $\mathcal{F}_r^{[-1]}$, $\mathcal{F}_r^{[0]}$ and $ \mathcal{F}_r^{[2]}$ are given in Eq.~(2.54), (2.56) and (2.59) of Ref.~\cite{Heffernan:2012vj} for the Kerr case, and $ \mathcal{F}_r^{[4]}$ is given in Eq.~(5.52) of Ref.~\cite{Heffernan:2012su} for the Schwarzschild case.

The regularized force in Eq.~(\ref{eq:regparams}) should include the monopole piece given in Eq.~(\ref{eq:monopole-force}).

\section{Implementation\label{sec:implementation}}

	\subsection{Numerics\label{sec:numerics}}
			\subsubsection{Homogeneous solution to the Teukolsky equations.}
In order to compute the components of the self-force, we need to evaluate radial Teukolsky functions $P^{lm}_{\pm 1}(r)$ and spin-weighted spheroidal harmonics $S_{\pm 1}^{lm}(\theta)$ at the particle's location, that is $r=r_0$ and $\theta = \pi/2$. To do so, we use the BlackHolePerturbation toolkit \cite{BHPToolkit}.
The angular functions are computed using the \textit{SpinWeightedSpheroidalHarmonics} package
and the radial functions are computed with the \textit{Teukolsky} package of the toolkit. The Teukolsky package implements the Mano-Suzuki-Takasugi (MST) method~\cite{Mano:1996vt,Mano:1996mf} to compute the homogeneous solution of the Teukolsky equations.

		\subsubsection{High-l tail contribution}

Our approach to compute the self-force requires us to sum over spin-weighted spheroidal modes or scalar spherical modes. 
Ideally one would sum an infinite number of modes but in practice we can only compute a finite number of components, up to $\ell_{\text{max}}$. 
In the case of the dissipative components of the self-force, the magnitude of the terms to be summed over decays exponentially, as can be seen in Fig~\ref{fig:multipoles}, and therefore the error from truncating the sum is negligible.
However, for the regularised conservative part of the self-force, the terms in the sum decay as an inverse power of $L = \ell + 1/2$ instead of an exponential, and the associated error from neglecting the higher modes is sizable. To reduce this error, we estimate the contribution coming from the $\ell > \ell_{\text{max}}$ modes following the standard approach of~\cite{Barack:2007tm} which we outline below.

In the large-$\ell$ regime, the modes of the regularized force in Eq.~(\ref{eq:Fr-reg}) are approximately
\begin{equation}
\mathcal{F}_r^{\text{reg}[n]\ell} \approx \frac{D_n}{L^{n}}, \label{eq:tail}
\end{equation}
where $n$ denotes the regularization order (with $n = 6$ in the Schwarzschild case and $n= 4$ in the Kerr case) and $D_n$ is a numerical coefficient to be determined by fitting to the high-$\ell$ modes.  Figure~\ref{fig:fr_schwa_reg} shows that Eq.~(\ref{eq:tail}) is a reasonable approximation for high values of $\ell$.
The contribution of the high-$\ell$ modes is then approximately
\begin{equation}
\sum_{\ell=\ell_{\text{max}}+1}^{\infty}\mathcal{F}_r^{\text{reg}[n]\ell} \approx \sum_{\ell=\ell_{\text{max}}+1}^{\infty} \frac{D_n}{L^{n}} = D_n \, \zeta(n,\ell_{\text{max}}+1),
\end{equation}
where $\zeta(s,a)$ is the Hurwitz Zeta function. 
	
	\subsubsection{Projection}
	
	In order to apply the mode-sum regularisation procedure, we need to project the force onto the scalar spherical harmonics basis. The original quantities in the spin-weighted spheroidal harmonics (associated to the index $l$) are first projected onto the spin-weighted spherical harmonics (associated with the index $\hat{l}$) which are then expanded onto scalar spherical harmonics (associated with the index $\ell$).
For the subdominant terms, which are proportional to $\cos \theta$ and $\cos^2 \theta$, one extra projection is needed (associated with the index $\ell_1$ and $\ell_2$).
Due to the presence of the 3j-symbols, and their association to spin-weighted or scalar quantities, the summation indices satisfy
\begin{eqnarray}
 \ell - 2 \ \leq \  &\ell_2 & \  \leq \ \ell + 2 \\
 \ell - 1 \ \leq \  &\ell_1 & \  \leq \ \ell + 1 \\
 \ell - (1+n) \ \leq \ &\hat{l} & \ \leq \ \ell + (1+n) \\
 |m| \ \leq \ & l&,\ \hat{l},\ \ell ,\ \ell_1 , \ \ell_2 
\end{eqnarray}
where $n = 0,1,2$ when computing the dominant, subdominant or subsubdominant term.

Since one spin-weighted spheroidal mode couples to several scalar spherical modes, we first compute all spin-weighted spheroidal modes separately and then perform the sums. 
We start by summing over $\ell_1$ or $\ell_2$ if we are computing the subdominant contributions at fixed $m$, $\hat{l}$ and $l$. 
We then sum over $m$ modes with fixed $\hat{l}$ and $l$ and then we sum over $\hat{l}$ modes at fixed $l$. 
All these sums performed at this point are finite and can be performed for any value of $l$. 
Finally we sum over $l$ which in principle can take any non-zero integer values. 
In practice however, we sum over a finite number of $l$ modes and estimate the contribution of the higher $l$ as described above.
	 
	 \subsection{Validation\label{sec:validation}}
	 
In order to validate our numerical code when computing the energy fluxes at infinity and through the horizon, we compare the total flux with the dissipative component of the self-force computed using \eqref{Ft_numerics}. We check that the two quantities agree up to numerical accuracy according to \eqref{conservation_law}. Furthermore, each flux is computed at $r=r_0^{+}$ and $r=r_0^{-}$ using different solutions to the homogeneous Teukolsky equations. We verify that the two fluxes obtained agree to numerical accuracy, meaning that our dissipative component of the self-force is continuous across the particle.
	 
	 In the case of the conservative piece of the self-force, we do not have a conservation law to support our numerical code. To validate our numerical approach in this case, we first verify that the radial component of the self-force is continuous across the particle as in the conservative component case. We note that while $\mathcal{F}_r$ is continuous across the particle, up to the expected precision, each spherical harmonic component $\mathcal{F}_r^{\ell}$ is discontinuous (for $a \neq 0$). We also observe that the sum of the even (odd) $\ell$ modes are independently continuous across the particle. Both features are likely due to the fact that we are only using a finite number of terms when expanding around $\cos\theta \approx 0$. 

We observe that the bare modes, $\mathcal{F}_r^\ell$, are well regularised using the regularisation parameters found in the literature~\cite{Heffernan:2012su,Heffernan:2012vj}. Finally, our result for the conservative self-force in the Schwarzschild case agrees with the results of Haas \cite{Haas:2011np} (see Fig.\ref{fig:deltas_schwa}).

\section{Results\label{sec:results}}
Below we present a selection of numerical results for the self-force. Where a dimensionless value is stated, e.g.~$\tilde{\mathcal{F}}_r$, the physical value should be inferred by reinstating the dimensionful constants, e.g. $\mathcal{F}_{r} = (q^2/4\pi \epsilon_0) (c^2 / GM)^2 \tilde{\mathcal{F}}_r$. 

 \subsection{Dissipative effects\label{sec:dis-results}}

  \subsubsection{Total fluxes}

\begin{figure}
 \includegraphics[width=10cm]{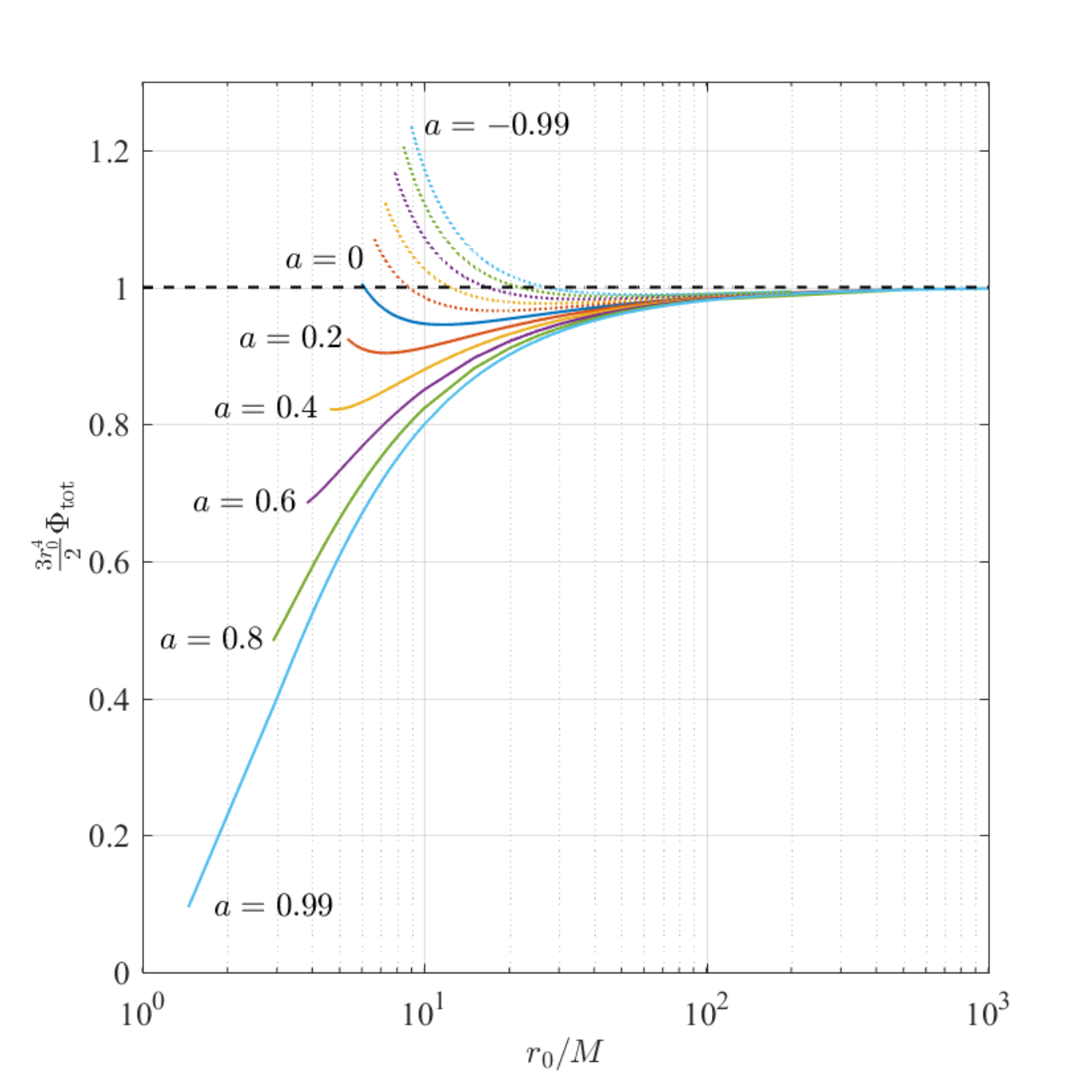}
 \caption{The radiated flux for an electromagnetically charged particle on a circular orbit at radius $r = r_0$ around a Kerr black hole of spin $a$. The flux $\Phi$ has been scaled by $3 r_0^4 / 2 q^2 G^2 M^2$ (see Eq.~(\ref{eq:Newtflux})). The solid lines correspond to prograde orbits ($a>0$), while the dotted lines correspond to retrograde orbits ($a<0$), and the color of the lines gives the magnitude of $a$. In each case, the minimum radius is the innermost stable circular orbit. 
 }
 \label{fig:totalflux}
\end{figure}

Figure \ref{fig:totalflux} shows the total energy flux $\Phi$ for a charged particle on a circular orbit about a black hole, as a function of orbital radius. The total flux is related to the self-force component $\mathcal{F}_t$ by Eq.~\eqref{conservation_law}. In the large-$r_0$ limit, the flux approaches an asymptotic value of $\Phi_{\text{Newt}}$, where (after restoring dimensionful constants)
\be
\Phi_{\text{Newt}} = \frac{2}{3} \frac{q^2}{4 \pi \epsilon_0 c^3} \frac{G^2 M^2}{r_0^{4}} .  \label{eq:Newtflux}
\ee
In Appendix \ref{appendix:Newtonian}, it is shown that $\Phi_{\text{Newt}}$ results from combining Keplerian orbits with the Abraham-Lorentz force (\ref{eq:abraham-lorentz}).

By fitting the numerical results in the weak-field region ($r_0 \gg M$), we infer that, for the flux at infinity, $\Phi_{\infty} \approx  \Phi_{\text{Newt}}$ at leading order, with a linear-in-$a$ contribution of $-\tfrac{8}{3} a r_0^{-11/2}$ at leading order. For the horizon flux, we infer that $\Phi_{h} \approx \tfrac{8}{3} r_0^{-7}$ at leading order for the Schwarzschild case, with a linear-in-$a$ contribution of $-\tfrac{2}{3} a r_0^{-11/2}$ at leading order in the Kerr case. Note that, for the horizon flux, the Kerr term begins at a \emph{lower} order in the expansion in $1/r_0$ than the Schwarzschild term.

\begin{figure}
 \subfigure[Scalar field $s=0$.]{\includegraphics[width=8.2cm]{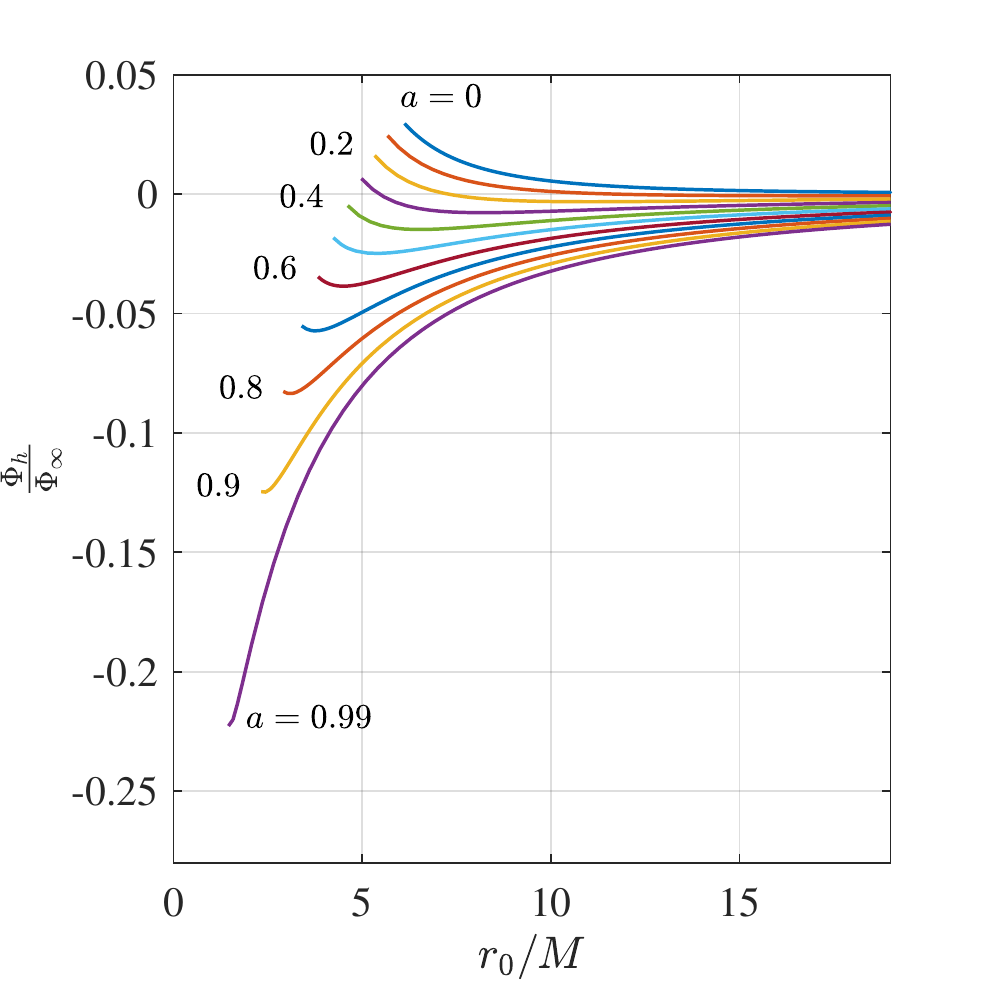}\label{fig:fluxratio_s0}}
 \subfigure[Electromagnetic field $s=1$.]{\includegraphics[width=8.2cm]{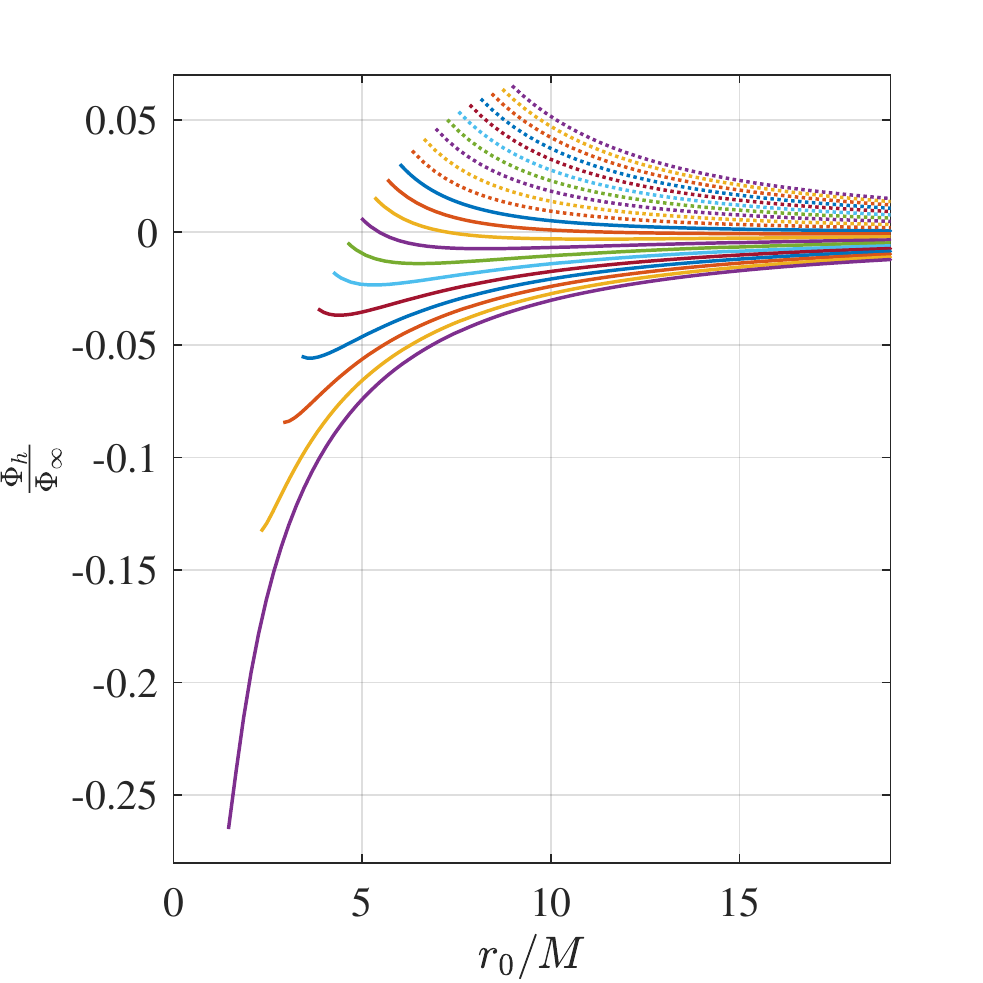}\label{fig:fluxratio_s1}}
 \subfigure[Gravitational field $s=2$.]{\includegraphics[width=8.2cm]{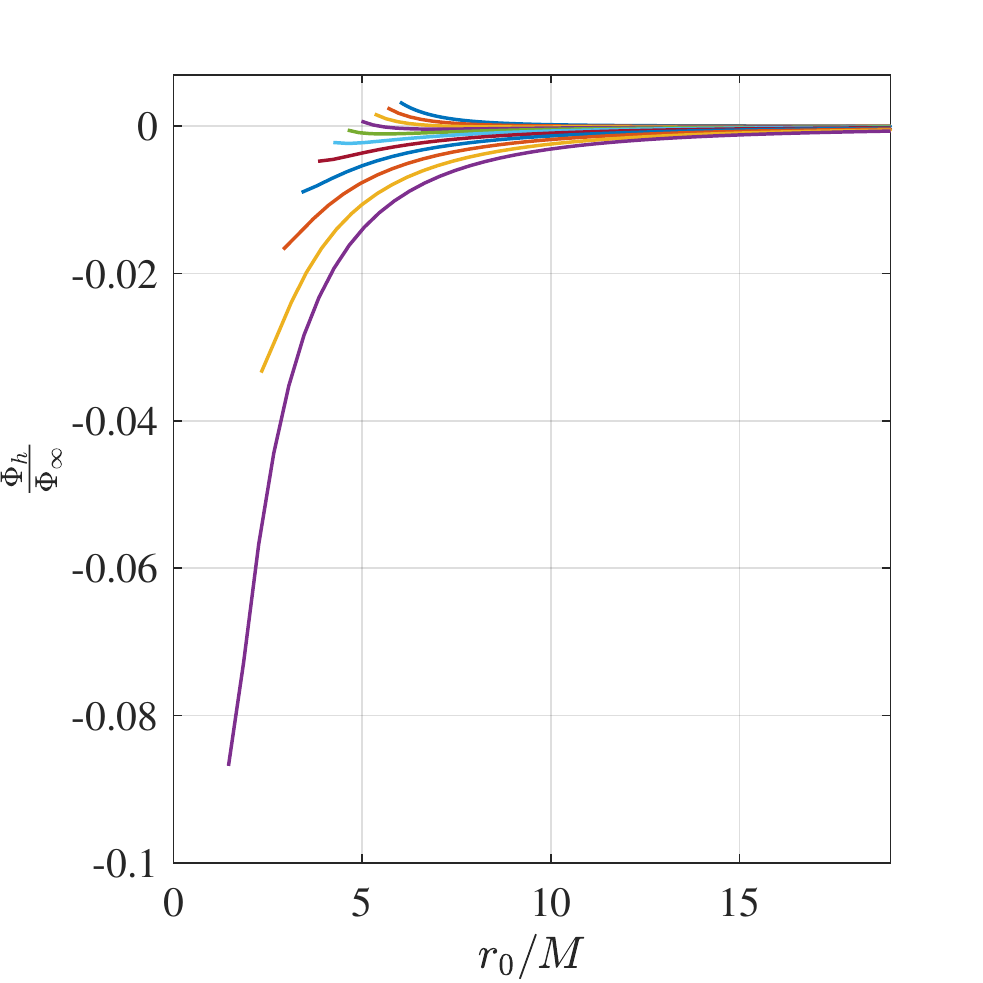}\label{fig:fluxratio_s2}}
 \caption{The ratio of the energy flux falling onto the horizon, $\Phi_h$, to the energy flux radiated to infinity, $\Phi_\infty$, as a function of orbital radius $r_0$, for various spin parameters $a$, and for the scalar, electromagnetic and gravitational cases. The solid lines correspond to prograde orbits ($a\geq 0)$. The dotted lines on the second plot, corresponds to retrograde orbits ($a<0$). The color of the lines gives the magnitude of $a$. Negative ratios arise due a negative flux from the horizon associated with superradiance. A value less than $-1$ would indicate the existence of floating orbits. } 
 \label{fig:fluxratios}
\end{figure}

Figure \ref{fig:fluxratios} shows the ratio of the flux through the horizon to the flux radiated away to infinity, for the three types of field (scalar, electromagnetic and gravitational). The scalar and electromagnetic cases are qualitatively similar, with radiation emitted principally in the dipole ($\ell=1$) modes. For particles that are orbiting in the same sense and the black hole spin, superradiance can lead to a significant extraction of energy from the horizon. For $a=0.99M$, the energy extracted from the hole is up to $\sim 26.5\%$ of that radiated away in the EM case, and up to $\sim 22.3 \%$ in the scalar-field case. Since this ratio falls below the threshold for balance ($100 \%$), there are no floating orbits. In the gravitational case, radiation is emitted principally in the quadrupole ($\ell = 2$) modes, and the maximum ratio is smaller ($\sim 8.7\%$ for $a=0.99M$). Again, there are no floating orbits. 

In the gravitational case, these results are consistent with those previously presented by Kapadia, Kennefick and Glampedakis~\cite{Kapadia:2013kf}. 

\begin{figure}
 \includegraphics[width=10cm]{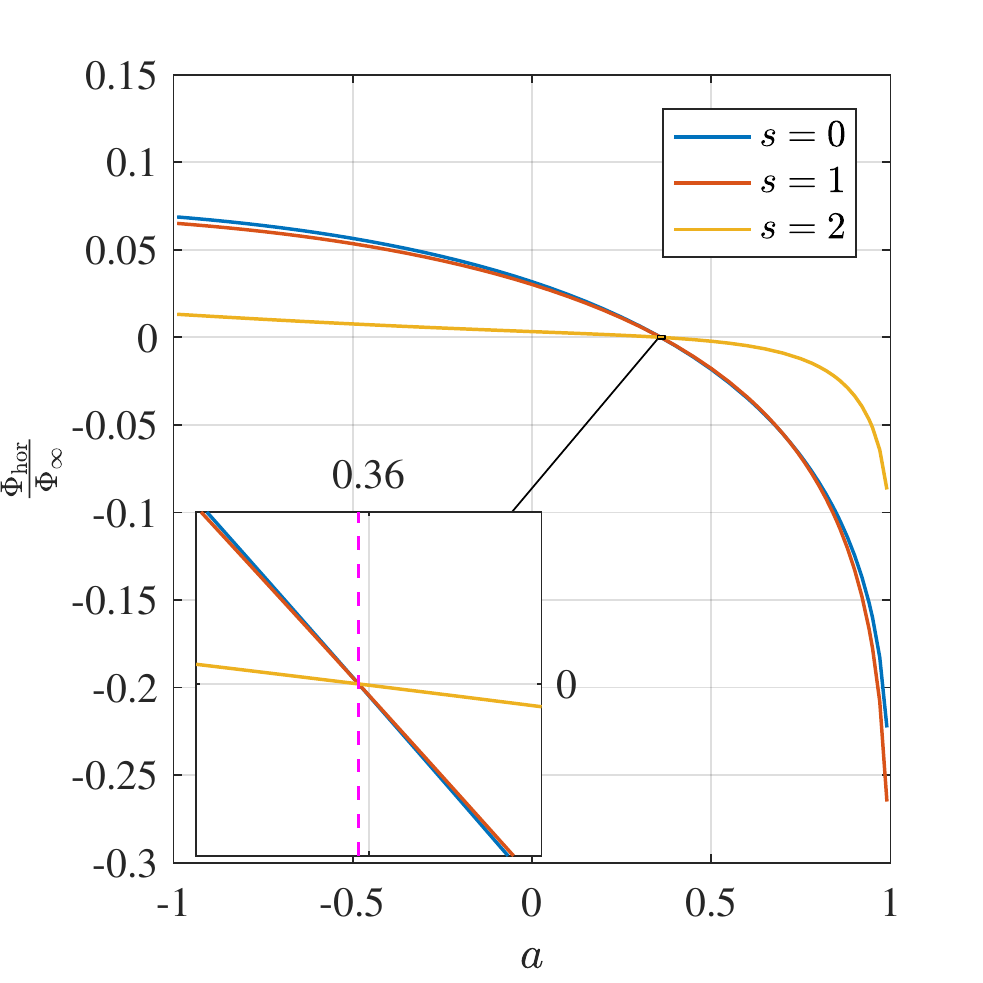}
 \caption{The ratio of fluxes $\Phi_h / \Phi_{\infty}$ for a particle on the innermost stable circular orbit, as a function of $a$ the spin of the black hole. Negative (positive) values of $a$ correspond to retrograde (prograde) circular orbits. For $\Omega_h > \Omega$, there is a negative flux ($\Phi_h < 0$) from the horizon, a manifestation of superradiance.}
 \label{fig:flux_at_isco}
\end{figure}

Figure \ref{fig:flux_at_isco} shows the ratio of fluxes $\Phi_h / \Phi_{\infty}$ for a particle on the innermost stable circular orbit (ISCO), as a function of the spin of the black hole. The ratio changes sign at $a = a_c \approx 0.359403$. This is the value of $a$ at which the angular frequency of the ISCO orbit (see Eq.~(\ref{eq:isco})) matches the angular frequency of the event horizon $\Omega_h$. For $a > a_c$, the (prograde) horizon frequency exceeds the orbital frequency. In this case, the electromagnetic field slows the rotation of the black hole, generating superradiance, leading to an extraction of flux from the event horizon and $\Phi_{h} / \Phi_{\infty} < 0$. 

\begin{figure}
 \includegraphics[width=10cm]{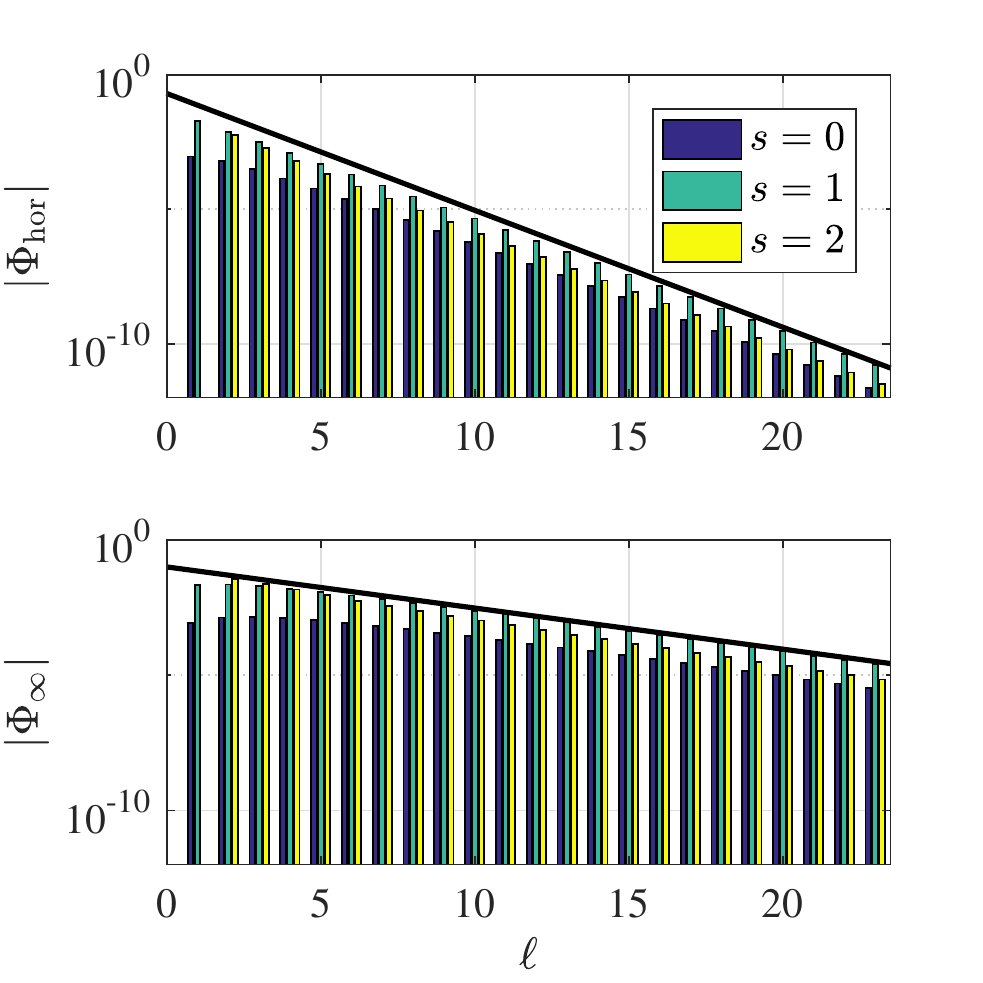}
 \caption{The multipolar structure of the flux radiated through the horizon (upper) and to infinity (lower) by a charged particle on a circular orbit at the ISCO of a Kerr black hole with $a = 0.99M$. The trendline indicates an exponential fall-off with multipole number $\ell$.}
 \label{fig:multipoles}
\end{figure}

Figure \ref{fig:multipoles} shows the multipolar structure of the flux generated by a particle at the ISCO for the scalar, electromagnetic and gravitational-wave cases. The lowest radiative multipole $\ell = \text{max}(|s|,1)$ generates the greatest flux at the horizon, and the low multipoles also dominate the flux at infinity. The plots show evidence for the expected exponential fall-off of the modal fluxes with $\ell+1/2$. 

 \subsection{Conservative effects\label{sec:con-results}}

	\subsubsection{Schwarzschild case}
	
\begin{figure}
 \includegraphics[width=10cm]{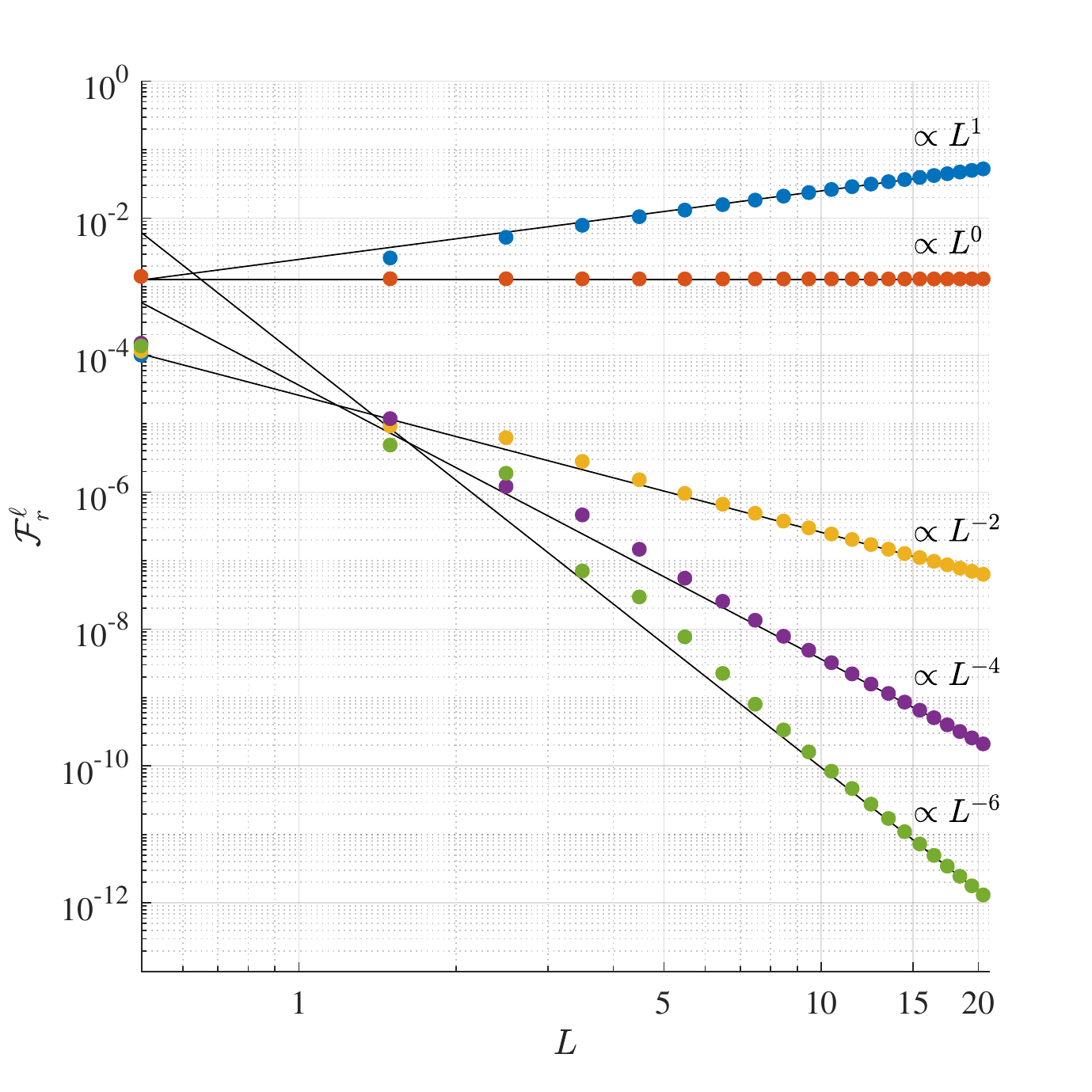}
 \caption{Scalar spherical modes of the radial component of the self-force $\mathcal{F}_r$, and regularisation at various orders. Here we have chosen $a=0$ (Schwarzschild) and $r_0/M = 20$. The blue dots are the values of the bare force, which grow linearly with $L = \ell + 1/2$ at large $L$. The solid black lines are guidelines to represent the decay of the regularised force.}
 \label{fig:fr_schwa_reg}
\end{figure}

{\it Regularisation.} Figure \ref{fig:fr_schwa_reg} illustrates the application of the regularization procedure to the radial component of the self-force, in the $a=0$ case. The unregularized (`bare') modes scale with $L = \ell + 1/2$ in the large-$\ell$ limit. After subtracting $\mathcal{F}^{[-1] \ell}_r$ and $\mathcal{F}^{[0] \ell}_r$ as in Eq.~(\ref{eq:regparams}), that is, removing the leading and subleading order regularization terms, one obtains modes that scale with $L^{-2}$. This is the minimum necessary to obtain a convergent sum. To reduce the error associated with the high-$\ell$ tail, and to demonstrate that our results match expectations, we removed a further two regularization terms, that is, we subtracted $\mathcal{F}^{[4] \ell}_r$, leaving a mode sum whose terms converge as $L^{-6}$ in the large-$\ell$ regime, as shown in Fig.~\ref{fig:fr_schwa_reg}.

{\it Weak field expansion}. Using numerical data for the radial component of the self-force at large values of $r_0$ we infer a weak-field expansion in the form
	\begin{equation}\label{eq:numerical_weakfield}
	\mathcal{F}_r(r_0) \approx \frac{q^2}{4 \pi \epsilon_0 c^2} \frac{GM}{r_0^3} \left( 1 + \frac{3}{2 \tilde{r}_0} + \frac{\alpha_2 \log(\tilde{r}_0)}{\tilde{r}_0^2} + \frac{\alpha_3}{\tilde{r}_0^2} + o\left( \frac{1}{\tilde{r}_0^3}\right) \right).
	\end{equation}
where $\tilde{r}_0 = r_0 / (GM/c^2)$. 
The coefficients $\alpha_2$ and $\alpha_3$ were estimated from summing over the first 15 $\ell$-modes, with data in two ranges (i) $1000 < r_0 < 1500$ and (ii) $900<r_0<1000$, yielding
\begin{subequations}
\begin{align}
\alpha_2 & = 1.249(2) &
\alpha_2 &= 1.231(1) \\
\alpha_3 &= 1.38(1) &
\alpha_3 &= 1.48(1).
\end{align}
\end{subequations}
The numeral in parantheses is the confidence interval in the final digit quoted, which is specific to the particular data set used for the fitting. The data supports the presence of a log term at sub-sub-leading order, but accurate estimates for $\alpha_2$ and $\alpha_3$ have not been obtained.

\begin{figure}
 \includegraphics[width=10cm]{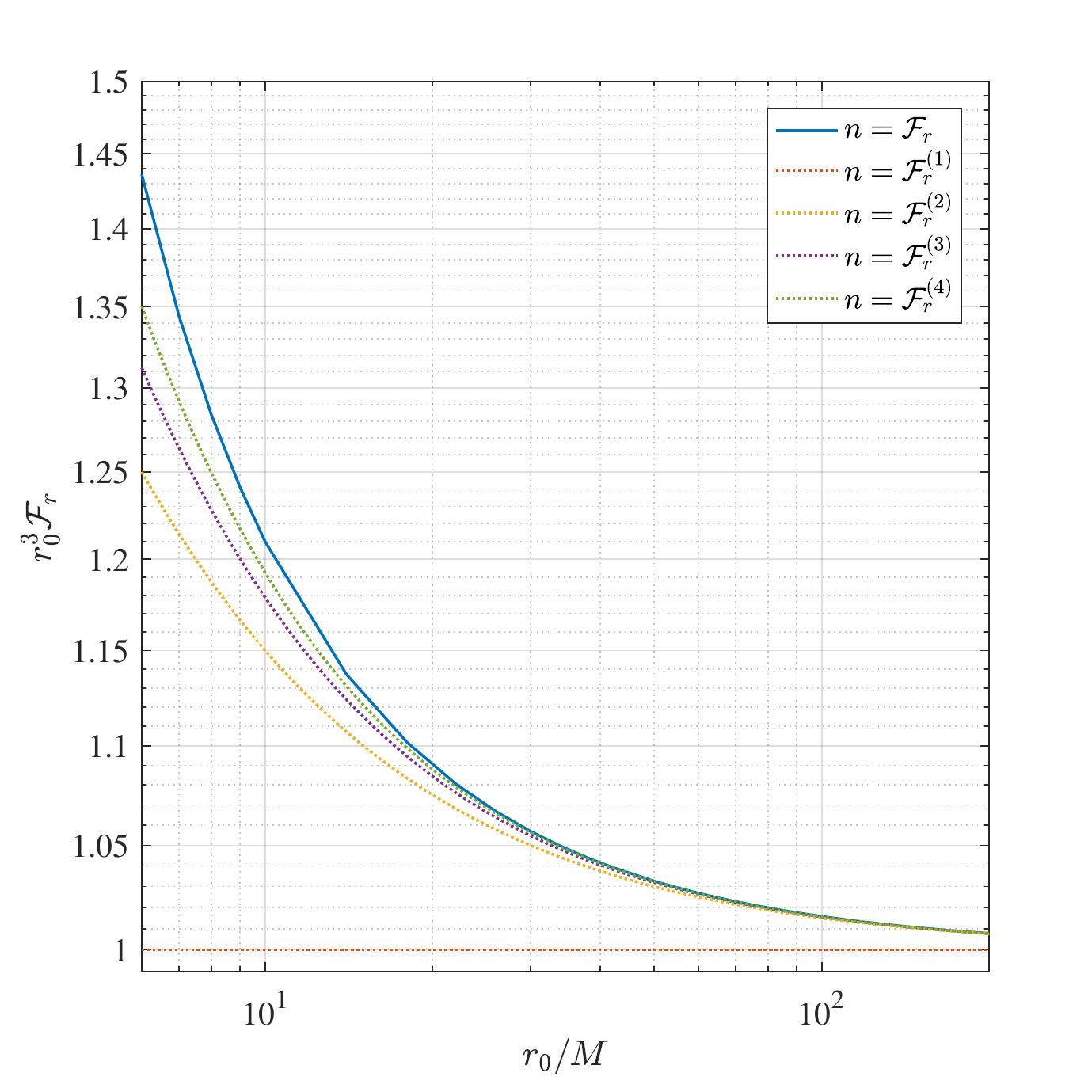}
 \caption{Comparison between the numerically-determined value of $\mathcal{F}_r$ (solid) and the weak-field expansion in Eq.~\eqref{eq:numerical_weakfield} (dashed), for the Schwarzschild ($a=0$) case. Here the dashed lines $\mathcal{F}_r^{(n)}$ show truncated versions of \eqref{eq:numerical_weakfield}, with $n$ indicating the number of terms included.  }
 \label{fig:fr_schwa_expansion}
\end{figure}

Figure \ref{fig:fr_schwa_expansion} compares the weak-field expansion, Eq.~(\ref{eq:numerical_weakfield}), with numerical data for $\mathcal{F}_r$ for $a=0$. It shows that $\mathcal{F}_r$ increases monotonically as $r_0$ decreases. Moreover, $\mathcal{F}_r$ differs from the leading order term in Eq.~(\ref{eq:numerical_weakfield}) by no more than a factor of $\sim 1.44$ across the range $[r_{\text{isco}}, \infty)$. Including successive terms in the expansion improves the agreement with the data; and Eq.~(\ref{eq:numerical_weakfield}) gives a relative error of $\sim 6\%$ at the ISCO. 

\begin{figure}
 \includegraphics[width=10cm]{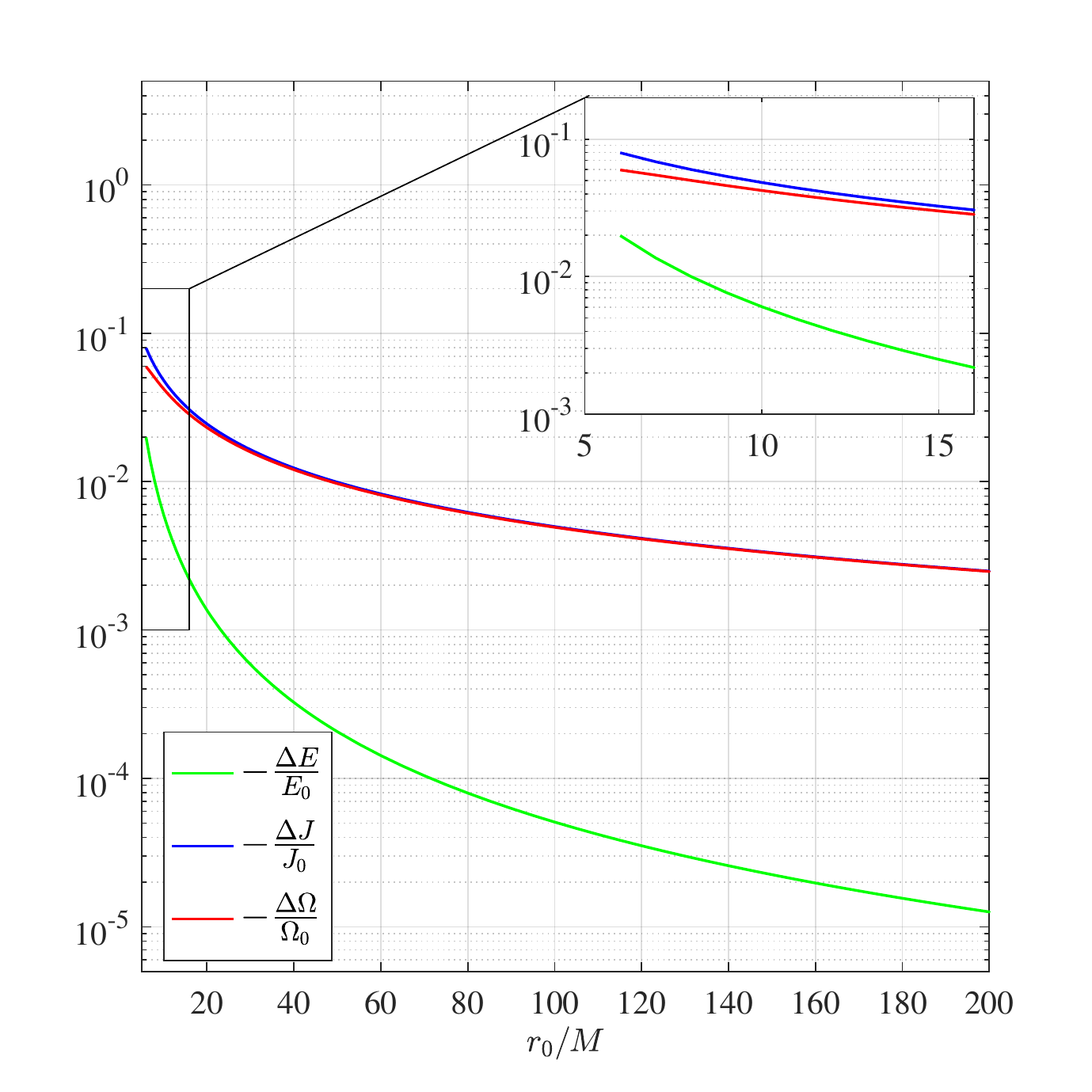}
 \caption{Fractional change of the energy (green), angular momentum (blue) and frequency (red) for a particle on a circular orbit as a function of the orbital radius $r_0$ in the Schwarzschild case. The inset shows the radial range presented in Fig.~12 of Haas \cite{Haas:2011np}. Our results agree with \cite{Haas:2011np} and provides the behaviour of the fractional for a larger radial range.}
 \label{fig:deltas_schwa}
\end{figure}

{\it Shifts in orbital parameters}.
The conservative self-force has the effect of shifting the orbital parameters from their geodesic values at order $q^2$. 
For circular orbit, the fractional change in the orbital energy $E$, angular momentum $J$ and frequency $\Omega$ is given by
\begin{subequations}
\begin{align}
\frac{\Delta\Omega}{\Omega_0} &= -\frac{(r_0 - 3M)r_0}{2\mu M}\mathcal{F}_r, \\
\frac{\Delta E}{E_0} &= -\frac{r_0}{2\mu} \mathcal{F}_r, \\
\frac{\Delta J}{J_0} &= - \frac{(r_0 - 2M)r_0}{2\mu M}\mathcal{F}_r.
\end{align}
\end{subequations}
Figure \ref{fig:deltas_schwa} shows the shift in $E$, $J$ and $\Omega$ as a function of $r_0$. In each case, the self-force leads to a reduction in $E$, $J$ and $\Omega$. The shifts for the Kerr case are given in Appendix \ref{appendix:shifts}. 

	\subsubsection{Kerr case}
\begin{figure}
 \includegraphics[width=10cm]{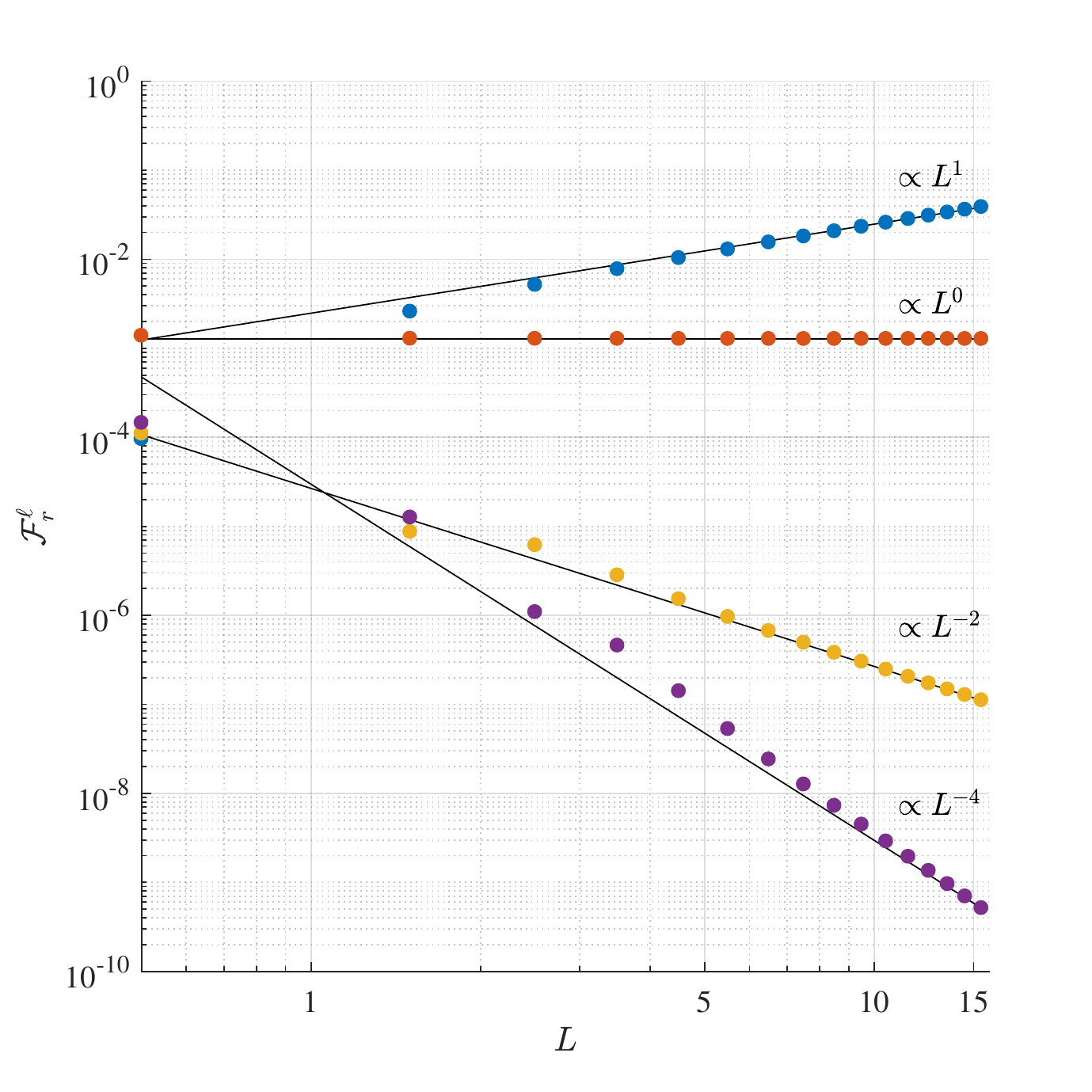}
 \caption{Scalar-spherical modes of the radial component of the self-force and their regularisation at various orders in the Kerr case. Here we have chosen $r_0/M = 20$ and $a = \tfrac{1}{2} M$. The blue dots are the values of the bare force, which (at leading order) grow linearly with $L = \ell + 1/2$. The solid black lines are guidelines to indicate the power-law decay of the regularised modes.}
 \label{fig:fr_kerr_reg}
\end{figure}

Figure \ref{fig:fr_kerr_reg} shows that the `bare' modes of the force, $\mathcal{F}_r^\ell$ defined in Eq.~(\ref{eq:Frl-kerr}), are correctly regularized with the regularization parameters calculated by Heffernan \emph{et al.} \cite{Heffernan:2012vj}. This is a non-trivial test of the formulation, and of the projection onto spherical harmonics. In the projection step, we find that it is necessary to expand to sub-sub-leading order in $z=\cos\theta$ in Eq.~(\ref{eq:Frl-kerr}) to achieve regularization at order $n=2$, and to obtain a regularized force $\mathcal{F}^{\text{reg}[2]}_r$ which is well-defined on the particle such that its left-sided limit ($r \rightarrow r_0^-$) and right-sided limit ($r \rightarrow r_0^+$) are in agreement.

\begin{figure}
 \includegraphics[width=10cm]{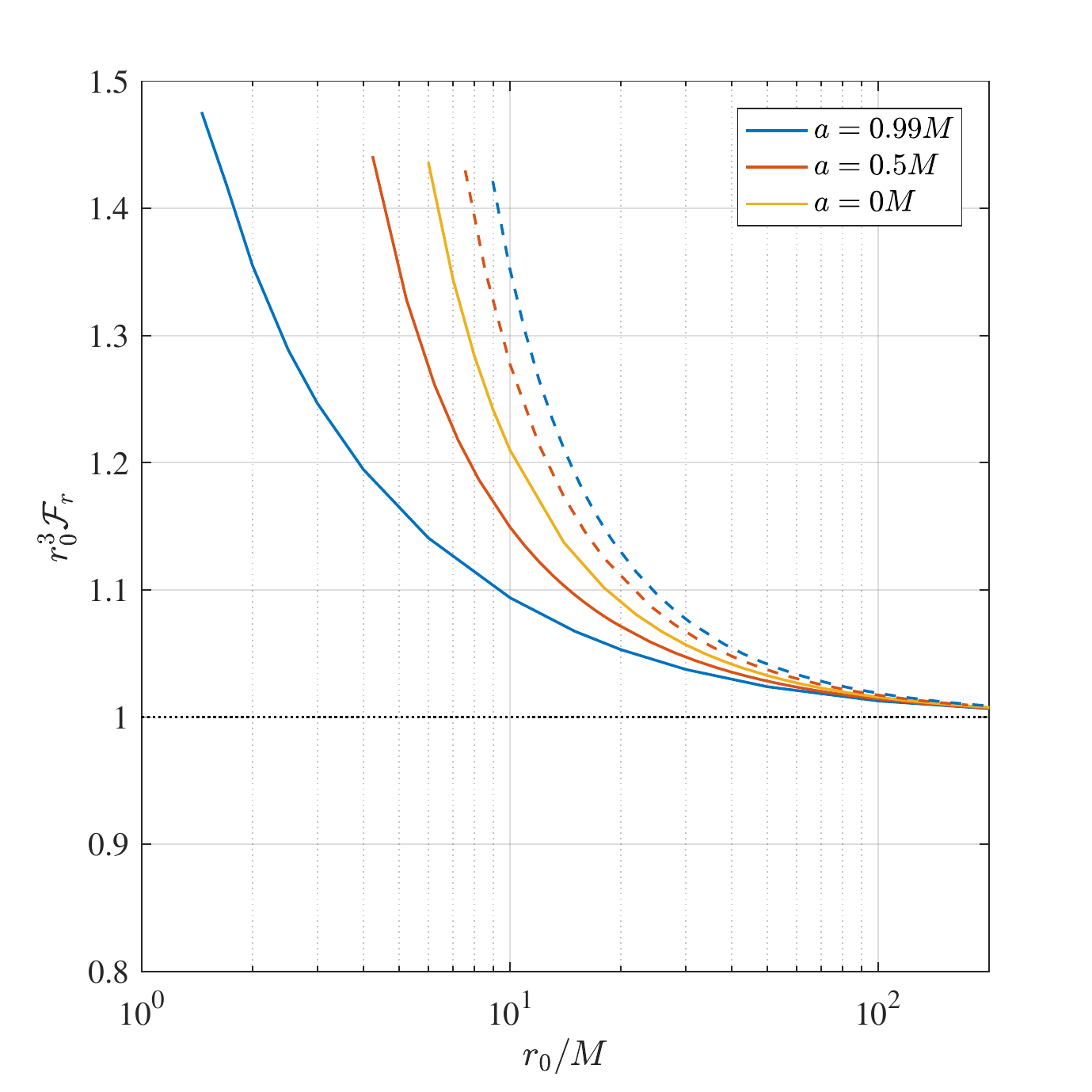}
 \caption{Radial component of the self-force (scaled by $r_0^3$) for various black hole spins, $a \in \{-0.99,0.5,0,0.5,0.99\}$. The solid lines correspond to prograde orbits ($a>0$), while the dotted lines correspond to retrograde orbits ($a<0$), and the color of the lines gives the magnitude of $a$. In each case, the minimum radius is the innermost stable circular orbit. }
 \label{fig:fr_vs_r_Kerr}
\end{figure}

Figure \ref{fig:fr_vs_r_Kerr} shows $\mathcal{F}_r$ as a function of $r_0$, for several values of the black hole spin parameter $a/M$. We observe that $\mathcal{F}_r$ is everywhere positive (i.e.~repulsive) and greater than $q^2/r_0^{3}$. At fixed radius, $\mathcal{F}_r$ is larger on the retrograde orbit than on the prograde orbit. The effect of black hole rotation increases as $r_0$ decreases, as expected. 

By fitting the numerical data, we find a linear-in-$a$ contribution to $\mathcal{F}_r$ of $-3 a r_0^{-9/2}$ at leading order.

\begin{figure}
 \includegraphics[width=10cm]{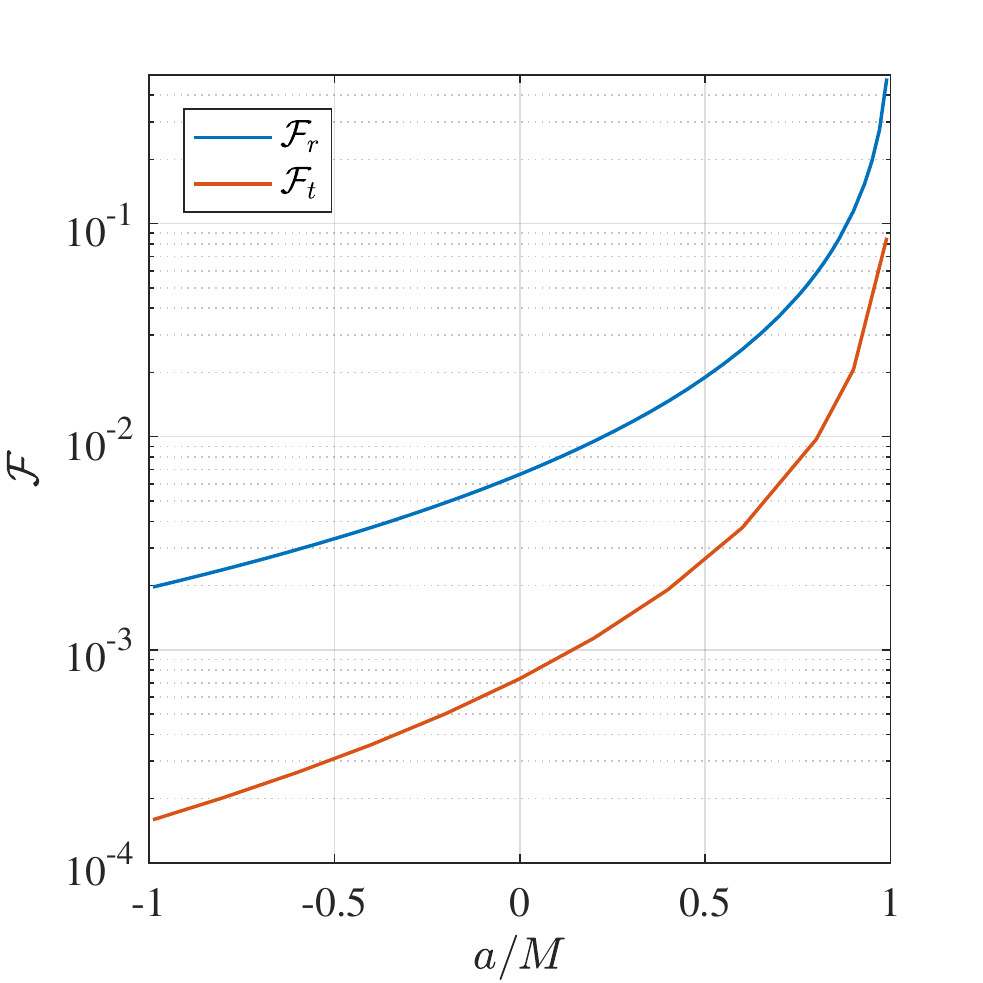}
 \caption{The radial and time components of the self-force for a particle on the innermost stable circular orbit, as a function of $a$ the spin of the black hole. Negative (positive) values of $a$ correspond to retrograde (prograde) circular orbits.}
 \label{fig:frvsa_isco}
\end{figure}

Figure \ref{fig:frvsa_isco} shows the self-force on the ISCO, as a function of $a/M$. The conservative component, $\mathcal{F}_r$, is always positive (i.e.~repulsive). The total flux is always positive, indicating that superradiance is insufficient for a floating orbit to arise. The magnitudes of $\mathcal{F}_r$ and $\mathcal{F}_t$ are largest on the corotating ISCO of a rapidly-rotating black hole. In the limit $a \rightarrow M$, the ISCO approaches $r_0 = M$. 

Table \ref{tbl:Fr} provides a selection of values of $\mathcal{F}_r$ for circular orbits of radii $r_0 \in [r_{\text{isco}}, 50M]$, for the black hole spin parameters $a = 0$, $\pm 0.5M$ and $\pm 0.99M$. 

\begin{table}
 \begin{tabular}{l || l | l | l | l | l} 
   & \multicolumn{5}{c}{$\mathcal{F}_r(r_0)$}                                                                                                         \\ \hline
 $r_0/M$& \multicolumn{1}{c|}{$a=-0.99$} & \multicolumn{1}{c|}{$a=-0.5$} & \multicolumn{1}{c|}{$a=0$} & \multicolumn{1}{c|}{$a=0.5$} & \multicolumn{1}{c}{$a=0.99$} \\ \hline
  $r_{\text{isco}}/M$ & 0.001967652(2)  &  0.003315094(1) & 0.0066497(5) & 0.019003(2) & 0.479(1)\\
  $10$ & 0.0013513595(1) & 0.0012770754(1) & 0.00120985(2) &0.00114927(1) & 0.001093823(1) \\
  $20$ & 0.000141150327(2) & 0.00013867449(5) & 0.00013624(1) & 0.000133916(2) & 0.0001316275(1)\\
  $50$ & $0.000008332378(2)$ & $0.000008296911(1)$ & $0.000008261044(2)$ & 0.000008225470(2) & 0.000008190833(6)
 \end{tabular}
 \caption{\emph{Radial component of self-force for circular equatorial geodesic orbits.} The dimensionless values in the table correspond to $\mathcal{F}_r / [(q^2/4\pi \epsilon_0) (c^2 / GM)^2]$. The digit in parantheses is an estimate of the uncertainty in the final quoted digit. The ISCO radius, defined in Eq.~(\ref{eq:isco}), is $r_{\text{isco}} / M \in \{ 8.971861, 7.554585, 6.0, 4.233003, 1.454498 \}$ (to 7 s.f.) for the cases $a/M \in \{-0.99, -0.5, 0, 0.5, 0.99 \}$.  }
 \label{tbl:Fr}
\end{table}

\section{Discussion and conclusion\label{sec:conclusion}}

In this article, we have computed the electromagnetic self-force acting on a point charge -- or, with caveats, on a charged compact body -- on a circular geodesic lying in the equatorial plane of a rotating black hole. This represents the first EM self-force calculation on Kerr spacetime in a dynamical scenario (see below for static cases). Our results complement those already available for the \emph{gravitational} self-force on Kerr \cite{Shah:2010bi,Shah:2012gu,Isoyama:2014mja, vandeMeent:2016hel,vandeMeent:2016pee, vandeMeent:2017bcc,vandeMeent:2018rms}, a topic which has received much attention due to its relevance in modelling Extreme Mass-Ratio Inspirals for gravitational wave detectors.

To compare the dissipative effects of the electromagnetic and gravitational self-forces, consider once more the inspiral of a particle or compact body of mass $\mu$ and charge $q$ into a black hole of mass $M$, driven by the dissipative component of the self-force. From the chirp formulae (\ref{eq:chirpEM}) and (\ref{eq:chirp}), valid in the large-$r_0$ regime, an order-of-magnitude estimate of the merger timescale, starting with an orbit of radius $r_0$, is
\begin{subequations}
\begin{align}
\tau_{\text{EM}} &\sim \left( \frac{\pi \epsilon_0 G M_{\odot}^2}{\mathcal{Q}^2} \right) \cdot \left( \frac{M}{\mu} \right) \cdot \left( \frac{r_0}{GM/c^2} \right)^2  \cdot \frac{r_0}{c} , \label{timescale-EM}\\
\tau_{\text{grav}} &\sim \quad \left( \frac{5}{2^{1/3} \, 32 } \right) \cdot \left( \frac{M}{\mu} \right) \cdot \left( \frac{r_0}{GM/c^2} \right)^3 \cdot \frac{r_0}{c} . \label{timescale-grav}
\end{align}
\end{subequations}
Here $\mathcal{Q}$ is the net charge density of the particle/compact body in Coulombs per solar mass, and we have made the assumption that $\mu \ll M$ to obtain (\ref{timescale-grav}). Numerical evaluation of the first parantheses in Eq.~(\ref{timescale-EM}) yields $7.4 \times 10^{39} \text{C}^2 / \mathcal{Q}^2$, and thus, for a compact body, an electromagnetically-driven inspiral is much slower than a gravitationally-driven inspiral, unless the compact body can support implausibly-high net charge densities of $\mathcal{Q} \gtrsim 10^{18} \text{C}$ per solar mass. On the other hand, for an elementary charged particle the converse is true, as $\mathcal{Q} \approx 1.9 \times 10^{38}\text{C}$ per solar mass for a proton, for instance. That is, for a charged elementary particle, the EM inspiral is more rapid and the gravitational wave flux is negligible; but nevertheless, the inspiral into a black hole is exceedingly slow due to the suppressing factor $M/\mu$. Of course, an elementary-particle-black-hole-inspiral scenario is rather artificial, not least because we have neglected all contents of the universe but two.

One key result of this work is a demonstration that the local dissipative component of the self-force $\mathcal{F}_t$ exactly balances with the sum of the electromagnetic flux radiated to infinity and down the horizon of the black hole, in accord with Eq.~\eqref{conservation_law}, up to the expected numerical precision. Closer examination of the fluxes, in Fig.~\ref{fig:totalflux}, \ref{fig:fluxratios} and \ref{fig:flux_at_isco}, shows that superradiance is stimulated when the angular velocity of the black hole horizon exceeds the orbital angular velocity. However, we find that superradiance is not sufficient to support floating orbits, even at the ISCO (see also \cite{Kapadia:2013kf}).

A key difference between the electromagnetic self-force and the gravitational self-force is that the latter is \emph{gauge-dependent} under small changes in the coordinate system at $O(\mu)$. More precisely, for circular orbits the dissipative component of the gravitational self-force -- relating to the radiated fluxes -- \emph{can} be identified uniquely, but the conservative component can not; it is coordinate-dependent. This means that it is \emph{not} possible to directly compare $\mathcal{F}_r$ between the electromagnetic and gravitational cases. Instead, one must look to the gauge-invariant consequences of the conservative component of self-force to make meaningful comparisons. For example, Fig.~\ref{fig:deltas_schwa} shows the fractional change in the orbital energy, angular momentum and frequency at fixed $r_0$ due to the conservative component of the self-force.

One such gauge-invariant consequence, slightly beyond the scope of this work, is the shift in the ISCO at $O(q^2)$ that arises due to the conservative component of the self-force. This can be calculated by examining mildly-eccentric orbits \cite{Barack:2009ey}, or possibly by using a Hamiltonian approach with circular-orbit data as input \cite{Isoyama:2014mja}; a comparison with known results for the ISCO shift induced by the gravitational self-force would certainly be of interest. Another observable that could be compared directly is the self-force-induced shift in the advance of the periapsis of an eccentric bound orbit  \cite{vandeMeent:2016hel}. 

The results presented in Sec.~\ref{sec:results} are numerical in nature, and we have inferred leading order terms in weak-field expansions by fitting the numerical data. A complementary approach is to apply the Mano-Suzuki-Takasugi (MST) formalism \cite{Mano:1996vt} to obtain analytical results in the form of high-order post-Newtonian expansions (see e.g.~\cite{Kavanagh:2015lva}). This has been done successfully in the gravitational self-force case, for quantities such as fluxes \cite{Fujita:2012cm,Fujita:2014eta,Munna:2020iju}, Detweiler's redshift invariant \cite{Bini:2015bla}, and the spin-precession invariant \cite{Bini:2015mza}. The MST method can be straightforwardly adapted from the $s=2$ to the $s=1$ case. An avenue for future work, therefore, is to apply the MST method to the formulae herein to obtain high-order expansions of (e.g.) $\mathcal{F}_t$ and $\mathcal{F}_r$ in closed form. 

It is worth noting that the calculation presented here is not fully self-consistent, in the sense that we have evaluated the self-force by assuming the past worldline of the particle is a \emph{geodesic}, rather than a trajectory that has itself been accelerated by its own self-force. Introducing the `true' trajectory would introduce sub-dominant contributions to the force starting at $O(q^4)$. One challenge, for future investigation, is to evolve the orbit in a fully self-consistent manner under the action of the electromagnetic self-force. This has already been done successfully for the gravitational self-force \cite{Warburton:2011fk,vandeMeent:2018rms}.

The electro\emph{static} self-force on a charged particle on Kerr was examined many years ago by L\'eaut\'e and Linet \cite{Leaute:1982}, and later by Piazzese and Rizzi \cite{Piazzese:1991}. For the special case of a particle at rest on the symmetry axis $\theta=0$ at $r=r_0$, the (repulsive, conservative) self-force is available in closed form \cite{Piazzese:1991}, 
\be
\mathcal{F}^\mu_{\text{self}} = \frac{q^2 (Mr_0 - a^2)}{(r_0^2+a^2)^2} e_3^\mu , \label{eq:Piazzese}
\ee
where $e_3^\mu$ is a unit spacelike vector along the symmetry axis. It is notable that Eq.~(\ref{eq:Piazzese}) does not depend on the sign of $a$, and thus frame-dragging effects are absent in this highly symmetric case. Here, we have established that $\mathcal{F}_r$ has a linear-in-$a$ contribution for geodesic orbits in the equatorial plane. 

Two further avenues of enquiry suggest themselves. First, the self-force on the ISCO in the $a \rightarrow M$ extremal limit has been investigated in the gravitational self-force context \cite{Gralla:2015rpa}, but not yet in the electromagnetic self-force context. Second, an additional physical effect which has not been examined here is the self-torque that would arise at $O(q^2)$ if the particle (or compact body) is endowed with a magnetic dipole moment. In other words, the force arising from the (regularized) magnetic field in the rest frame of the particle.

\begin{acknowledgments}
With thanks to Barry Wardell and Niels Warburton for email correspondence and discussions. This work makes use of the Black Hole Perturbation Toolkit \cite{BHPToolkit}. T.T.~and S.D.~acknowledge financial support from the Science and Technology Facilities Council (STFC) under Grant No.~ST/P000800/1. S.D.~acknowledges financial support from the European Union's Horizon 2020 research and innovation programme under the H2020-MSCA-RISE-2017 Grant No.~FunFiCO-777740.  
\end{acknowledgments}

\appendix

\section{Dissipative self-force in the Newtonian limit\label{appendix:Newtonian}}
For circular orbits far from a black hole ($r_0 \gg GM/c^2$), the speed of the particle, $|v| = r_0 \Omega = \sqrt{GM/r_0}$, is small in comparison with the speed of light $c$, and a leading-order Newtonian approximation for the flux (\ref{eq:Newtflux}) and the chirp formula (\ref{eq:chirpEM}) is obtained by combining the Abraham-Lorentz force (\ref{eq:abraham-lorentz}) with circular orbits in Newtonian gravity.

The work done in unit time $P$ upon a particle of charge $q$ by the Abraham-Lorentz force (\ref{eq:abraham-lorentz}) is
\be
P = \mathbf{F} \cdot \mathbf{v} = \frac{2}{3} \frac{q^2}{4 \pi \epsilon_0 c^3} \, \dot{\mathbf{a}} \cdot \mathbf{v} .
\ee
Inserting a fixed circular orbit with $\mathbf{r} = r_0 \, \hat{\mathbf{r}}$ and $\mathbf{v} \equiv \dot{\mathbf{r}} = r_0 \Omega \hat{\boldsymbol{\phi}}$ and $\mathbf{a} \equiv \dot{\mathbf{v}} = -r_0 \Omega^2 \hat{\mathbf{r}}$ and $\dot{\mathbf{a}} \equiv  -r_0 \Omega^3 \hat{\boldsymbol{\phi}}$, where $\hat{\mathbf{r}}$ and $\hat{\boldsymbol{\phi}}$ are unit vectors and $\Omega = \sqrt{GM/r_0^3}$ is the angular frequency of the orbit, yields
\begin{align}
P &= - \frac{2}{3} \frac{q^2}{4 \pi \epsilon_0 c^3} r_0^2 \Omega^4 \\
 &=  - \frac{2}{3} \frac{q^2}{4 \pi \epsilon_0 c^3} \frac{G^2 M^2}{r_0^4} .
\end{align}
By conservation of energy, the flux radiated to infinity is equal and opposite to the work done on the particle by the Abraham-Lorentz force, that is, $\Phi_{\text{Newt}} = -P$, yielding Eq.~(\ref{eq:Newtflux}).

For a particle of mass $\mu$ on a circular orbit under gravity, the sum of kinetic and (Newtonian) potential energies is $E = - G M \mu / 2 r_0$. We now allow the particle to gradually spiral inwards on a sequence of quasi-circular orbits, by equating $P$ with $\dot{E} = -GM\mu \dot{r} / 2 r^2$. This leads to
\be
\frac{\dot{f}}{f^3} = \frac{8 \pi^2 q^2}{4 \pi \epsilon_0 c^3},  
\ee
where $f = \Omega / 2 \pi$ is the orbital frequency. Integrating with respect to time leads to
\be
f(t) = \frac{1}{4 \pi} \sqrt{\frac{4 \pi \epsilon_0 c^3 \mu}{q^2}} \, (t_0 - t)^{-1/2} , \label{eq:chirp2}
\ee
where the time of collision $t_0$ arises as the constant of integration. For the case of an electron of mass $\mu = m_e$ and charge $q=-e$, Eq.~(\ref{eq:chirp2}) reduces to Eq.~(\ref{eq:chirpEM}) once we insert the definition of the fine-structure constant $\alpha = e^2 / (4 \pi \epsilon_0 \hbar c)$ and the Bohr radius $a_0 = 4 \pi \epsilon_0 \hbar^2 / (m_e e^2)$.

\section{Energy flux}\label{A:flux}

	\subsection{Flux at infinity}
At infinity the energy flux is given by
\begin{equation}
\Phi^{K}_R = \Delta t ^{-1}\int T^{ab}K_b d\Sigma_a
\end{equation}
where we have chosen $K^{\mu}=[1,0,0,0]$ to be the Killing vector and $d\Sigma_{\mu}$ is defined by the condition $r\rightarrow \infty $ and is given by
\begin{equation}
d\Sigma_\mu = n_\mu d\Sigma,
\end{equation}
with
\begin{equation}
n_\mu = \frac{[0,1,0,0]}{\sqrt{g^{rr}}} \quad \text{and} \quad d\Sigma = |h|^{1/2} dt d\theta d\phi.
\end{equation}
where $h_{\mu\nu}$ is the induced metric on the hypersurface define by $r\rightarrow \infty $. Since $|h| = g/g_{rr}$, we have
\begin{equation}
d\Sigma_\mu = [0,1,0,0] \ \sqrt{-g}  dtd\theta d\phi.
\end{equation}
The flux is then given by
\begin{equation}
\Phi_R = \int{T^r_{\ t} \sqrt{-g}  d\theta d\phi}.
\end{equation}
The energy-momentum tensor can be expressed in terms of the Maxwell scalars as~\cite{Teukolsky:1973ha}
\begin{eqnarray}
4\pi T_{\mu\nu} = & - \left\lbrace \phi_0 \phi_0^* n_\mu n_\nu + 2\phi_1 \phi^*_1[l_{(\mu} n_{\nu)} + m_{(\mu}m^*_{\nu)}] + \phi_2 \phi_2^*l_\mu l_\nu \right. \nonumber \\
&- \left. 4\phi_0^*\phi_1 n_{(\mu}m_{\nu)} - 4\phi_1^* \phi_2l_{(\mu}m_{\nu)} + 2\phi_2\phi_0^* m_{\mu}m_\nu \right\rbrace + c.c.
\end{eqnarray}
where parentheses denote symmetrization. With our choice of tetrad, we find that the relevant terms as $r\rightarrow \infty$ are 
\begin{equation}
\lim_{r \rightarrow \infty} T^{r}_{\ t} = \frac{1}{2\pi}\left(  \phi_2 \phi_2^* - \frac{\phi_0\phi_0^*}{4}\right)
\end{equation}
We recall that 
\begin{eqnarray}\label{phi_decomposition}
\phi_0 &=& \sum_{l,m} {}_{+1}R_{lm}(r) \Splus^{\ell m}(\theta)e^{im(\phi - \Omega t)},\\
 2(r - ia\cos\theta)^2\phi_2 &=& \sum_{l,m} {}_{-1}R_{lm}(r) \Sminus^{\ell m}(\theta) e^{im(\phi - \Omega t)},
\end{eqnarray}
as well as the fact that
\begin{equation}
\lim_{r\rightarrow\infty} r^2 |{}_{+1}R_{lm}(r)| = 0 \quad \text{and} \quad \lim_{r\rightarrow\infty}  \frac{|{}_{-1}R_{lm}(r)|}{r} = |{}_{-1}\alpha_{lm}^{\infty}|.
\end{equation}
Using the orthonormality properties of the spin-weighted spheroidal harmonics, we therefore get the energy flux radiated at infinity
\begin{equation}
\Phi_\infty = \lim_{r\rightarrow \infty} \int{\frac{r^2\sin\theta}{2\pi} \phi_2 \phi_2^*} d\theta d\phi= \sum_{l,m}\frac{|{}_{-1}\alpha_{lm}^{\infty}|^2}{8\pi}
\end{equation}		

	\subsection{Flux through the horizon}
	
In order to evaluate the flux of energy through the horizon, we first need to modify the tetrad basis we have defined in eq.~\eqref{eq:tetrad1} since it is singular at the horizon. We first perform a rotation of class III (according to Chandrashekar's convention):
\begin{equation}
l^\mu \rightarrow \frac{\Delta}{2(r^2 + a^2)} l^\mu = \left[ \frac{1}{2}, \frac{\Delta}{2(r^2 + a^2)},0,\frac{a}{2(r^2 + a^2)}\right]
\end{equation}
and
\begin{equation}
n^\mu \rightarrow \frac{2(r^2 + a^2)}{\Delta}n^\mu = \left[\frac{(r^2 + a^2)^2}{\Sigma\Delta},-\frac{(r^2 + a^2)}{\Sigma},0,a\frac{(r^2 + a^2)}{\Sigma\Delta} \right].
\end{equation}
We then go to a Kerr-Schild frame via the coordinate transformation:
\begin{equation}
dv = dt + \frac{r^2 + a^2}{\Delta}dr \quad \text{and} \quad d\tilde{\phi} = d\phi + \frac{a}{\Delta}dr.
\end{equation}
In this frame, the null vectors $l^{(HH)}$ and $n^{(HH)}$, where $HH$ stands for Hartle-Hawking, are given by
\begin{eqnarray}
l^{(HH)\ \mu} &=& \left[ 1,\frac{\Delta}{2(r^2 + a^2)},0,\frac{a}{r^2 + a^2}\right], \\
n^{(HH)\ \mu} &=& \left[ 0, - \frac{r^2 + a^2}{\Sigma},0,0\right].
\end{eqnarray}
It is important to note that on the horizon, the vector $l^{(HH)}$ can be expressed in terms of the time and angular Killing vector $K_T^{\mu} = [1,0,0,0]$ and $K_L^{\mu} = [0,0,0,1]$
\begin{equation}\label{l2Killing}
l^{(HH)} = K_T + \Omega_h K_L.
\end{equation}
where $\Omega_h = a/2Mr_{+}$ is the angular frequency of the horizon. 
In this basis, which is well behaved at the horizon, the Maxwell scalars $\phi_0^{(HH)}$ is related to the Maxwell scalar $\phi_0$ computed in the basis~\eqref{eq:tetrad1} via
\begin{equation}
\phi_0^{(HH)} = \frac{\Delta}{2(r^2+a^2)}\phi_0.
\end{equation}
The surface element $d\Sigma_a$ of the horizon (which is a null hypersurface) is given by
\begin{equation}
d\Sigma_\mu = l^{(HH)}_\mu d\sigma dt
\end{equation}
where $d\sigma = 2Mr_{+}\sin(\theta)d\theta d\phi$ is the elementary surface area of the event horizon.
Therefore, the elementary flow of energy and angular momentum through the horizon are
\begin{eqnarray}
\left( \frac{d^2\Phi_{h}^{T}}{dtd\Omega}\right) &=& 2Mr_{+}l^{(HH)}_{\mu} K_{T}^{\nu} T^{\mu}_{\ \nu} \\
\left( \frac{d^2\Phi_{h}^{L}}{dtd\Omega}\right) &=& 2Mr_{+}l^{(HH)}_{\mu} K_{L}^{\nu} T^{\mu}_{\ \nu} 
\end{eqnarray}
Combining these with \eqref{l2Killing} and using the fact that $\Phi^T = - \Omega \Phi^{L}$ we get
\begin{equation}
\left( \frac{d^2\Phi_{h}^{T}}{dtd\Omega}\right) = \frac{2Mr_{+}\omega}{\omega - m\Omega_h}T^{\mu\nu}l^{(HH)}_\mu l^{(HH)}_\nu.
\end{equation}
By definition $T^{\mu\nu}l^{(HH)}_\mu l^{(HH)}_\nu = \phi_0^{(HH)}\left.\phi_0^*\right.^{(HH)}/2\pi$, therefore we finally have
\begin{equation}
\left( \frac{d^2\Phi_{h}^{T}}{dtd\Omega}\right) = \frac{\omega}{8Mr_{+}\tilde{\omega}}\frac{\Delta}{2\pi} \phi_0\phi_0^*.
\end{equation}
Integrating over the surface element using the decomposition \eqref{phi_decomposition}, the orthonomality of the spin-weighted spheroidal harmonics, and the asymptotic behaviour of the radial function near the horizon, we obtain
\begin{equation}
\Phi^{T}_h = \sum_{l,m}{\frac{\omega}{16\pi Mr_{+}\tilde{\omega}} |{}_{+1}\alpha_{lm}^{h}|^2}.
\end{equation}

\section{Projection onto scalar spherical harmonics\label{A:projection}}
To compute the physical conservative part of the self-force we apply the mode-sum regularisation procedure. As a preliminary step before applying the regularization, one should decompose the radial force onto a basis of scalar spherical harmonics. Since the structure of the Kerr metric invited us to use spin-weighted spheroidal harmonics as a basis for the angular functions of our problem, we now need to project the spin-weighted spheroidal harmonics onto scalar spherical harmonics.

\subsection{Projection of the spin-weighted spheroidal harmonics}

\subsubsection{From spin-weighted spheroidal harmonics to spin-weighted spherical harmonics}
We first decompose the spin-weighted spheroidal harmonics $S_s^{lm}$ onto the spin-weighted spherical harmonics $Y_s^{lm}$:
\begin{equation}
S_s^{lm}(\theta) = \sum_{\hat{l}} \left(b_s^m \right)^l_{\hat{l}} \ Y_s^{\hat{l}m}(\cos \theta).
\end{equation}
The coefficients $\left(b_s^m \right)^l_{\hat{l}}$ are computed using the Black Hole Perturbation Toolkit \cite{BHPToolkit}.

\subsubsection{From spin-weighted spherical harmonics to scalar spherical harmonics}
We decompose the spin-weighted spherical harmonics in terms of spherical harmonics $Y^{lm}_0$,
\begin{eqnarray}
Y_{+1}^{lm}(z) &=& \sum_{\tilde{l}} \frac{(A_{+1}^{m})_{\tilde{l}}^{l} }{\sqrt{1-z^2}} Y_0^{\tilde{l}m}(z),\\
Y_0^{lm}(z) &=& \sum_{\tilde{l}} \delta^{l}_{\tilde{l}} \, Y_0^{\tilde{l}m}(z), \\
Y_{-1}^{lm}(z) &=& \sum_{\tilde{l}} \frac{(A_{-1}^m)_{\tilde{l}}^{l}  }{\sqrt{1-z^2}}Y_0^{\tilde{l}m}(z),
\end{eqnarray}
where $z=\cos\theta$ and the coefficients are given by
\begin{subequations}
\begin{eqnarray}
(A_{+1}^{m})_{\tilde{l}}^{l} &=& (-1)^{m+1}\sqrt{2(2l+1)(2\tilde{l}+1)}\begin{pmatrix}
1 & l & \tilde{l} \\
0 & m & -m 
\end{pmatrix}
\begin{pmatrix}
1 & l & \tilde{l} \\
1 & -1 & 0 
\end{pmatrix} , \\
(A_{-1}^{m})_{\tilde{l}}^{l} &=& (-1)^{m} \  \sqrt{2(2l+1)(2\tilde{l}+1)}\begin{pmatrix}
1 & l & \tilde{l} \\
0 & m & -m 
\end{pmatrix}
\begin{pmatrix}
1 & l & \tilde{l} \\
-1 & 1 & 0 
\end{pmatrix} .
\end{eqnarray}
\label{eq:Aco}
\end{subequations}
It follows from the properties of the Wigner 3j symbols that
\begin{equation}
(A_{+1}^{m})_{\tilde{l}}^{l} = (-1)^{l + \tilde{l}} \, (A_{-1}^{m})_{\tilde{l}}^{l} . 
\end{equation}
Combining the two decompositions, we can write the spin-weighted spheroidal harmonics as
\begin{equation}
S_s^{lm}(\theta) = \sum_{\hat{l},\tilde{l}} \left(b_{s}^m\right)^{l}_{\hat{l}} \left(A_{s}^{m}\right)_{\tilde{l}}^{\hat{l}} \frac{Y_0^{\tilde{l}m}(z)}{\sqrt{1-z^2}}.
\end{equation}
Note that due to the presence of the 3j-symbols in Eqs.~(\ref{eq:Aco}), the indices $\tilde{l}$ and $\hat{l}$ satisfy $\tilde{l} - 1 \leq \hat{l} \leq \tilde{l} + 1$.

\subsection{Expansion of $\mathcal{L}_1 S_{+1}^{lm}(\theta)$}
The definition of $\mathcal{L}_1 \Splus^{l m}$ is
\begin{equation}
\mathcal{L}_1 \Splus^{l m}(\theta) = \partial_\theta  \Splus^{lm}(\theta) + \left(\frac{m}{\sin\theta} - a\omega \sin\theta \right) \Splus^{lm} + \frac{\cos\theta}{\sin\theta} \Splus^{lm}. 
\end{equation}
In order to project $\mathcal{L}_1 \Splus^{\ell m}$ onto scalar spherical harmonics, we first need to project $\partial_\theta \Splus^{\ell m}$.

The spherical harmonics of different spins are related by
\begin{eqnarray}\label{L1def}
\eth \, Y_s^{lm} &=& \sqrt{(l-s)(l+s+1)} \, Y_{s+1}^{lm}, \\
\bar{\eth} \, Y_s^{lm} &=& -\sqrt{(l+s)(l-s+1)} \, Y_{s-1}^{lm},
\end{eqnarray}
where the spin-raising and spin-lowering operators $ \eth$ and $\bar{\eth}$ are defined by
\begin{eqnarray}\label{eth_rel}
\eth f_s (\theta,\phi) &=& -(\sin\theta)^s \left[\frac{\partial}{\partial \theta } + \frac{i}{\sin\theta}\frac{\partial}{ \partial \phi} \right]((\sin\theta)^{-s} f_s) \\
\bar{\eth} f_s (\theta,\phi) &=& -(\sin\theta)^{-s} \left[\frac{\partial}{\partial \theta } - \frac{i}{\sin\theta}\frac{\partial}{ \partial \phi} \right]((\sin\theta)^{s} f_s).
\end{eqnarray}
We have that
\begin{equation}
\partial_\theta  \Splus^{lm}(\theta)  = \sum_{\hat{l}} \left( b_{+1}^m \right)^{l}_{\hat{l}} \partial_\theta \, Y_{+1}^{\hat{l}m}(\cos \theta )  .
\end{equation}
We can eliminate the derivative using the relation \eqref{eth_rel} and the expression for $\bar{\eth}$, namely, 
\begin{eqnarray}
\bar{\eth} \, Y_{+1}^{lm}(z) &=& -(\sin\theta)^{-1} \left[\frac{\partial}{\partial \theta } - \frac{i}{\sin\theta}\frac{\partial}{ \partial \phi} \right]( \sin\theta \, Y_{+1}^{lm}(z)) \\
&=& -\frac{1}{\sin\theta} \left[ \left( \cos\theta + m\right) Y_{+1}^{lm}(z) + \sin\theta \, \partial_\theta Y_{+1}^{lm}(z) \right] \\
&=& -\sqrt{l(l+1)} \, Y_0^{lm}.
\end{eqnarray}
Therefore, we have that
\begin{equation}\label{partial_proj}
\partial_\theta  S_{+1}^{lm}(\theta) = \sum_{\hat{l}} \left( b_{+1}^m\right)^{l}_{\hat{l}} \left( \sqrt{\hat{l}(\hat{l}+1)}\ Y_0^{\hat{l}m}(\cos\theta) - \frac{\left( \cos\theta + m\right)}{\sin\theta} Y_{+1}^{\hat{l}m}(\cos\theta) \right).
\end{equation}
Substituting \eqref{partial_proj} into \eqref{L1def}, we get
\begin{eqnarray}
\mathcal{L}_1  S_{+1}^{lm} 
&=& \sum_{\hat{l}} \left( b_{+1}^m\right)^{l}_{\hat{l}} \left[ \sqrt{\hat{l}(\hat{l} +1)}\ Y_0^{\hat{l}m} - a m \Omega \sin\theta Y_{+1}^{\hat{l}m} \right] \\
&=&
\sum_{\hat{l},\tilde{l}} \left( b_{+1}^m \right)^{l}_{\hat{l}} \left[ \sqrt{\hat{l}(\hat{l} +1)} \, \delta^{\hat{l}}_{\tilde{l}} - a m \Omega  (A_{+1}^{m})_{\tilde{l}}^{\hat{l}} \right] Y_0^{\tilde{l}m}.
\end{eqnarray}

\section{Shifts in orbital parameters from conservative self-force\label{appendix:shifts}}

On Kerr spacetime, the shifts in the energy and angular momentum at fixed $r_0$ are
\begin{subequations}
\begin{align}
\frac{\Delta E}{E_0} &= - \frac{r_0  \Delta_0 (L_0 (r_0 - 2M) + 2aM E_0)}{2 E_0 \mathcal{X} } \mathcal{F}_r / \mu, \\
\frac{\Delta L}{L_0} &= - \frac{r_0 \Delta_0 \left( E_0 (r_0^3 + a^2(r_0 + 2M)) - 2 a M L_0 \right)}{2 L_0 \mathcal{X} }    \mathcal{F}_r / \mu , \\
\mathcal{X} &= (r_0^3 - 3 M r_0^2 - 2 M a^2) E_0 L_0 + a M \left( (3 r_0^2 + a^2) E_0^2 + L_0^2 \right)
\end{align}
\end{subequations}
and the shift in the angular velocity at fixed $r_0$ is
\begin{align}
\frac{\Delta\Omega}{\Omega_0} 
 &= 
 - \frac{\Delta_0^2 r_0^3 \left( (r_0^3 + a^2(r_0+2M)) E_0^2 - 4 E_0 L_0 a M - L_0^2 (r_0 - 2M) \right)}{2 \mathcal{X} (E_0 r_0 (r_0^2 + a^2) - 2 a M (L_0 - a E_0)) (r_0 L_0 - 2 M (L_0 - a E_0))}  \mathcal{F}_r/\mu .
\end{align}

\bibliographystyle{apsrev4-1}
\bibliography{refs}

\end{document}